\documentstyle[prl,aps,amssymb,epsf,psfig]{revtex}
\newcommand \beq{\begin{eqnarray}}
\newcommand \eeq{\end{eqnarray}}
\newcommand \be{\begin{eqnarray}}
\newcommand \ee{\end{eqnarray}}
\newcommand{\set}[2]{\newcommand{#1}{#2}}
\set{\pa}{\partial \over \partial\, }
\set{\leftvector}{\stackrel{\leftarrow}{\partial }}
\set{\rightvector}{\stackrel{\rightarrow}{\partial }}

\begin{document}
\twocolumn[\hsize\textwidth\columnwidth\hsize
           \csname @twocolumnfalse\endcsname
\title{Retarded versus time-nonlocal quantum kinetic equations}
\author{Klaus Morawetz}
\address{
Max-Planck-Institute for the Physics of Complex Systems, 
Noethnitzer Str. 38, 01187 Dresden, Germany}
\author{Pavel Lipavsk\'y, V\'aclav \v Spi\v cka }
\address{Institute of Physics, Academy of Sciences, Cukrovarnick\'a 10,
16200 Praha 6, Czech Republic}
\maketitle
\begin{abstract}
The finite duration of the collisions in Fermionic systems 
as expressed by the retardation
time in non-Markovian Levinson-type kinetic equations is discussed in the quasiclassical limit.
We separate individual contributions included in the memory effect 
resulting in (i) off-shell tails
of the Wigner distribution, (ii) renormalization of scattering rates
and (iii) of the single-particle energy, (iv) collision delay and (v)
related non-local corrections to the scattering integral. In this way
we transform the Levinson equation into the Landau-Silin equation 
extended by the non-local corrections known from the theory of dense
gases.
The derived nonlocal kinetic equation
unifies the Landau theory of quasiparticle transport with the
classical
kinetic theory of dense gases. The space-time symmetry is discussed
versus particle-hole symmetry and a solution is proposed which
transforms these two exclusive pictures into each other.
\end{abstract}
\vskip2pc]

\section{Introduction}
The generalization of the Boltzmann equation (BE) towards dense interacting 
quantum systems is a still demanding and unsolved task. In dense systems, the 
mean time between successive collisions becomes comparable to the 
collision duration. In this case all particles cannot be described as a system 
of weakly interacting quasiparticles but a set of particles bound in 
two - particle correlations has to be accounted for in a distinct manner. 
In the spirit of chemical reactions, one can introduce the quasiparticle
density, $n_f$ (a density of single-atom molecules), and the density of
particles in the correlated states, $n_c$ (twice the density of bi-atomic
molecules). The total density $n$ is their sum
\begin{equation}
n=n_f+n_c.
\label{nfc}
\end{equation}
In parallel with the partial pressure of molecules, ${\cal P}=k_BT(n_f+
{1\over 2}n_c)$, and the Guldberg-Waage law of mass action law, $n_c=Kn_f^2$,
even a small fraction of the correlated density, $n_c\ll n_f$, contributes 
to the second virial coefficient, ${\cal P}\approx k_BT\left(n-{1\over 2}K
n_f^2\right)$. The correlated density thus plays an important role in the 
thermodynamic behavior of the system.

While the equation of state of the ideal gas, ${\cal P}=k_BTn$, does not
reflect any microscopic properties of particles in the system, the second 
virial coefficient directly follows from the particle-particle interaction. 
In the early time of statistical mechanics, the experimental second virial 
coefficient has been used to deduce interaction potentials. Surprisingly,
the firmly established concept of the equilibrium virial expansion has 
never found a corresponding position in the theory of non-equilibrium 
systems  although a number of attempts have been made to modify the 
BE so that its equilibrium limit will cover at least the second virial 
coefficient \cite{B69,Ba84,S91}. The achieved corrections to the BE have 
a form of gradient or nonlocal contributions to the scattering
integral. 
For a hard-sphere gas the non-local correction is trivial, therefore
the main theoretical focus was on the statistical correlations
\cite{E21,CC90,HCB64,Sch91,C62,W63,W63a,W65,KO65,DC67,GF67,BE79}.
It turned out that the treatment of higher order contributions is far
from trivial as the dynamical statistical correlations result in
divergences that are cured only after a re-summation of an infinite set
of contributions. Naturally, this re-summation leads to non-analytic
density corrections to the scattering integral
\cite{W65,KO65,DC67,GF67,BE79}. The moral of these hard-sphere studies
is that beyond the non-local corrections one must sum up an infinite
set of contributions; the plain expansion leads to incorrect results.
From this point of view, the approximate statistical virial
corrections implemented by \cite{BGM94}, although possibly reasonable,
are not sufficiently justified by the theory.

Real particles do not interact like hard spheres, in particular when
their de Broglie wave lengths are comparable with the potential range.
An effort to describe the virial corrections for more realistic systems
resulted in various generalizations of Enskog's equation
\cite{B69,S91,W57,W58,W60,S60,S64,TS70,SS71,RS76,B75,M89,La89,TNL89,NTL89,L90,L90a,H90,H90a,H91,LM90,S90,NTL91,NTL91a,S95,SML95}.
By closer inspection one finds that all tractable quantum theories deal
exclusively with the non-local corrections. The statistical correlations
in quantum systems would require an adequate solution of three-particle
collisions (say from Fadeev equations) which are now intensively being studied
\cite{BRS96}. A systematic incorporation of three-particle
collisions into the kinetic equation, however, is not yet fully
understood, therefore we discuss only binary processes.

An alternative way to describe the correlation has been developed for 
Fermi systems at very low temperatures. Since the Pauli principle 
excludes all but the zero-angle scattering channels, the correlation can 
be re-cast into a renormalization of the single-particle energies known 
as the quasiparticle energies. While quasiparticles behave in many aspects 
like free particles, the density dependence of the quasiparticle energies 
results in non-trivial thermodynamic properties. In accordance with the 
focus either on zero-angle or finite-angle scattering, in the literature 
one can find two distinct classes of quantum kinetic equations. 

The zero-angle class are equations determining the time evolution of 
the momentum and space dependent quasiparticle distribution function 
$f(k,r,t)$. This leads to kinetic equations of the Landau-Silin type
\begin{equation}
{\partial f\over\partial t}+{\partial\varepsilon\over\partial k}
{\partial f\over\partial r}-{\partial\varepsilon\over\partial r}
{\partial f\over\partial k}=z((1-f)\sigma^<_\varepsilon -
f\sigma^>_\varepsilon).
\label{tr1}
\end{equation}
Here the drift term is characterized by the quasiparticle energies 
$\varepsilon(k,r,t)$ and the collision integral consists of scattering-in 
and -out terms which are summarized in the selfenergy as functionals 
of the quasiparticle distribution function, $\sigma^{\gtrless}[f]$. 
The collision rates are reduced by the wave function renormalization, 
which is given by the frequency derivative of the real part of the 
selfenergy $\sigma$ at the pole value $z^{-1}=1-{\partial\sigma\over
\partial\omega}|_{\omega=\varepsilon}$. This kinetic equation is local 
in time and space, i.e., it describes the collisions as instantaneous 
point--like events. Similarly to the BE, the local collisions yield no virial 
corrections. The virial corrections are thus covered exclusively by the 
quasiparticle energy while no correlated density appears.

The finite-angle class are equations which include the finite duration 
of collisions, i.e., they have non-Markovian scattering integrals. These 
kinetic equations are usually developed for the reduced density matrix 
or the Wigner function $\rho(k,r,t)$.  The Wigner distribution obeys 
a Levinson-type of kinetic equation \cite{L65,L69}
\begin{eqnarray}
{\partial\rho_t\over\partial t}&+&{\partial\epsilon^{\rm hf}\over\partial k}
{\partial\rho_t\over\partial r}-{\partial\epsilon^{\rm hf}\over\partial r}
{\partial\rho_t\over\partial k}
\nonumber\\
&=&2\,{\rm Im}\!\int\limits_{-\infty}^t\! d\bar t\, G^R_{t,\bar t}\left[
(1-\rho_{\bar t})\Sigma^<_{\bar t,t}-\rho_{\bar t}\Sigma^>_{\bar t,t}
\right].
\label{levinson}
\end{eqnarray}
Here the drift term is essentially determined by the Hartree-Fock mean 
field, $\epsilon^{\rm hf}=k^2/2m+\sigma_{\rm hf}$, and the double-time 
selfenergies $\Sigma^{\gtrless}_{\bar t,t}$ are given as functionals of 
the Wigner distribution. Explicit time arguments of the collision integral, 
$\bar t<t$, show the retardation which requires to include the propagation, 
$G^R_{t,\bar t}$, from the retarded time $\bar t$ to the time $t$ of 
balancing changes. Having the non-Markovian form, this kinetic equation 
is capable to describe finite duration effects and corresponding virial 
corrections. 

The Landau-Silin and Levinson equations have common features
with the original BE. In particular, they express the balance between
the drift given by the gradient terms on the left hand side and the 
dissipation given by the scattering integral on the right hand side.
On the other hand, there is a remarkable difference in the actual physical
contents of these common ingredients, namely $\rho\not=f$, $\varepsilon\not=
\epsilon^{\rm hf}$, and the right hand sides of (\ref{tr1}) and 
(\ref{levinson}) are rather different.

In the present paper we argue that the differences between the phenomenological
quasiparticle picture, Eq.~(\ref{tr1}), and the
BBGKY-type\footnote{Equations derived from the
  Born-Bogoliubov-Green-Kirkwood-Yvon (BBGKY) hierarchy of reduced densities.} equation
(\ref{levinson}) can be characterized in terms of the retardation of the 
collision integral of (\ref{levinson}). We will show that there are
various contributions to the total retardation, each responsible for 
different features seen in the quasiparticle picture extended by non-local
corrections.

If both approaches are equivalent, the Levinson equation (\ref{levinson}) 
must contain terms on the collisional side which should be possible to
rearrange into gradients such that the quasiparticle energies of the drift 
side in the Landau-Silin equation (\ref{tr1}) are obtained. The mutual 
relation between the Wigner distribution $\rho$ and the quasiparticle
distribution $f$ is of essential importance for such a rearrangement.

In Sec.~II we show that a part of the retardation of the Levinson equation 
(\ref{levinson}) describes the off-shell motion of particles. This 
off-shell motion can be eliminated from the kinetic equation which
requires to introduce an effective distribution (the quasiparticle
distribution $f$) from which the Wigner distribution $\rho$ can be
constructed (\ref{tr2})
\begin{equation}
\rho=f+\int{d\omega\over 2\pi} {\wp\over\omega-\varepsilon}
{\partial\over\partial\omega}
\left((1-f)\sigma^<_\omega-f\sigma^>_\omega\right).
\label{tr2a}
\end{equation}
This relation is the extended quasiparticle picture derived for small
scattering rates. 
The limit of small scattering rates has been first introduced by Craig
\cite{C66a}. An inverse functional $f[\rho]$ has been constructed in 
\cite{BD68}. For equilibrium non-ideal plasmas this approximation has been 
employed by \cite{SZ79,KKL84} and under the name of the generalized
Beth-Uhlenbeck approach has been used by \cite{SR87} in nuclear matter 
for studies of the correlated density. The authors in \cite{KM93} have 
used this approximation with the name extended quasiparticle 
approximation for the study of the mean removal energy and high-momenta 
tails of Wigner's distribution. The non-equilibrium form has been derived 
finally as the modified Kadanoff and Baym ansatz \cite{SL95}. We
will call it extended quasiparticle approximation.

We show that the retardation is responsible for
gradient terms by which the mean-field drift of the Levinson equation
(\ref{levinson}) differs from the quasiparticle drift of the 
Landau-Silin equation
(\ref{tr1}). In Sec.~III we discuss the remaining parts of the retardation,
a piece which is compensated by the decay of internal propagators, and
a piece which describes the collision delay. Special emphasize is put
on internal double counts of correlations in the Levinson
equation. Related to this question we present in appendix~C a
discussion of a common met pitfall in literature. In Sec.~IV we complete 
the kinetic equation with the finite collision delay and provide related
non-local corrections in space. Our main result is the nonlocal
kinetic equation derived here in an alternative way to 
\cite{SLM96,LSM97},
\begin{eqnarray}
&&{\partial f_a\over\partial t}+{\partial\varepsilon_a\over\partial k}
{\partial f_a\over\partial r}-{\partial\varepsilon_a\over\partial r}
{\partial f_a\over\partial k}
\nonumber\\
&&=\sum\limits_b \int {\cal P}\biggl \{
  f_a'f_b'(1-f_a)(1-f_b)-f_af_b(1-f_a')(1-f_b') \biggr \}
\nonumber\\&&
\label{e11ni}
\end{eqnarray}
with $f_a'=f_a(k\!-\!q\!-\!\Delta_K,r\!-\!\Delta_3,t\!-\!\Delta_t)$,
$f_b'=f_b(p\!+\!q\!-\!\Delta_K,r\!-\!\Delta_4,t\!-\!\Delta_t)$,
$f_a=f_a(k,r,t)$ and $f_b=f_b(p,r\!-\!\Delta_2,t)$.
The differential cross section ${\cal P}\sim |t^R|^2$ 
is proportional to the square of the
amplitude of the T-matrix.
All non-local corrections are given by derivatives of the scattering
phase shift \mbox{$\phi={\rm Im\ ln}t^R(\omega,k,p,q,t,r)$}
\cite{SLM96},
\begin{eqnarray}
\Delta_t&=&{\partial\phi\over\partial\omega},\ \ \
\Delta_E=-{1\over 2}{\partial\phi\over\partial t},\ \ \
\Delta_K={1\over 2}{\partial\phi\over\partial r},\ \ \
\Delta_3=-{\partial\phi\over\partial k},
\nonumber\\
\Delta_2&=&{\partial\phi\over\partial p}-{\partial\phi\over\partial q}-
{\partial\phi\over\partial k},\ \ \ \ \ \ \ \ \ \
\Delta_4=-{\partial\phi\over\partial k}-{\partial\phi\over\partial q}.
\label{e6}
\end{eqnarray}
We discuss in Sec.~IV the space-time symmetry of this
collision integral and link the correlated density with the collision
delay. In Sec.~V we summarize and refer to numerical solutions of the
derived nonlocal kinetic equation. The most important formulae for the
T-matrix are presented in appendix A and the technique how to derive
nonlocal corrections from gradient expansion is summarized in appendix
B.

\section{Kinetic equations}
First we want to show that the Levinson equation can be transformed into the kinetic equation of 
Landau-Silin type and vice versa.  Therefore we
first derive the kinetic equation for the Wigner distribution from 
the real-time Green function technique. For a homogeneous system similar 
investigations have been presented by Bornath at al \cite{BKKS96}. Here we
extend their approach to inhomogeneous systems and want to focus
specially on the physical contents of both equations. This approach
will furnish us with a set of Kramers-Kronig relations needed to link the Wigner 
and the quasiparticle distributions.

We consider the two independent correlation functions for Fermionic 
operators $G^>(1,2)=\langle a(1)a^+(2)\rangle$ and $G^<(1,2)=\langle 
a^+(2)a(1)\rangle$, where cumulative variables mean time and space,
$1\equiv t,x$. The time diagonal part of $G^<$ yields the Wigner
distribution, $\rho(x_1,x_2,t)=G^<(1,2)_{t_{1,2}=t}$. For the correlation 
function $G^<$ the Kadanoff and Baym equation of motion reads \cite{KB62}
\begin{eqnarray}
-i\left(G_0^{-1}G^<-G^<G_0^{-1}\right)=
&-&i\left(\Sigma^R G^<-G^<\Sigma^A\right)
\nonumber\\
&+&i\left(G^R \Sigma^<-\Sigma^<G^A\right).
\label{kb}
\end{eqnarray}
The retarded and advanced functions are introduced as $B^{R}(1,2)=-i
\theta(t_1-t_2)(B^>+B^<)$ and $B^{A}(1,2)=i\theta(t_2-t_1)(B^>+B^<)$. 
The products are understood as integrations over intermediate variables 
(time and space). The Hartree-Fock drift term has the form
\begin{eqnarray}
G_0^{-1}(1,2)&=&\left(i{\partial\over\partial t_1}+{\nabla_1^2\over 2 m} 
\right)\delta(1-2)
\nonumber\\
&-&\Sigma^{\rm hf}(x_1,x_2,t_1)\delta(t_1-t_2).
\label{hf}
\end{eqnarray}

The Kadanoff and Baym equation (\ref{kb}) is more general than any of the 
above kinetic equations which can be derived from it with the help of 
specific approximations. One possibility is to rearrange the terms such 
that they yield the Landau-Silin kinetic equation for the quasiparticle 
distribution function (\ref{tr2}) when the gradient expansion is applied 
\cite{bm90,SL94,SL95}. The other possibility is to collect the terms 
on the right hand side of (\ref{kb}) into a non-Markovian collision 
integral which results in the Levinson equation (\ref{levinson}) 
for the Wigner distribution 
\cite{JW84,LSV86,MR95}. The connection between both equations, however, has 
remained dubious because in the Levinson equation the drift term is 
controlled exclusively by the Hartree-Fock selfenergy while in the 
Landau-Silin equation the drift term is represented by the full 
quasiparticle energy. In the following we demonstrate how the 
Levinson equation can be converted into the Landau-Silin equation. To 
this end we first derive the Levinson equation from (\ref{kb}). 

\subsection{Precursor of the Levinson equation}
On the time diagonal, $t_{1,2}=t$, the Kadanoff and Baym equation 
(\ref{kb}) yields an identity
\begin{eqnarray}
-i\bigl(G_0^{-1}G^<-&&G^<G_0^{-1}\bigr)(t,x_1,t,x_2)
\nonumber\\
=\int\limits_{-\infty}^{t}\!dt'\int\! dx'&&
\Bigl(G^>(t,x_1,t',x')\Sigma^<(t',x',t,x_2)
\nonumber\\
&&+\Sigma^<(t,x_1,t',x') G^>(t',x',t,x_2)\Bigr)
\nonumber\\
-\int\limits_{-\infty}^{t}\!dt'\int\! dx'&&
\Bigl(G^<(t,x_1,t',x')\Sigma^>(t',x',t,x_2)
\nonumber\\
&&+\Sigma^>(t,x_1,t',x') G^<(t',x',t,x_2)\Bigr). 
\label{levin}
\end{eqnarray}
On the left hand side there is the Hartree-Fock drift and the right hand 
side contains a non-Markovian collision integral. Note that correlations 
beyond the Hartree-Fock field are exclusively in the collision integral.

\subsubsection{The Wigner representation}
Now we concentrate on the gradient expansion of the collision integral. 
The quasi-classical approximation is achieved using the Wigner mixed 
representation for space arguments
\begin{eqnarray} 
&&\hat\sigma\left(k,{x_1+x_2\over 2},{t_1+t_2\over 2},t_1-t_2\right)=
\nonumber\\ 
&& \ \ \ \ \ \ \ \
\int d (x_1\!-\!x_2) {\rm e}^{-ik(x_1-x_2)}\Sigma(t_1,x_1,t_2,x_2).
\label{sW}
\end{eqnarray}
As a rule, we use the lower case to denote operators in the full Wigner 
representation (momentum, space, time, frequency) and the hat to denote the
Wigner representation in space and double-time representation. Note that
time arguments of $\hat\sigma$ now denote the center-of-mass time, 
${t_1+t_2\over 2}$, and the difference time, $t_1-t_2$. These two Wigner
representations are identical for $\rho$ and $\sigma^{\rm hf}$ which do
not depend on the difference time.

In the representation (\ref{sW}) the identity (\ref{levin}) reads
\begin{eqnarray}
&&\left({\pa t}+\frac{k}{m}{\pa r} \right )\rho(k,r,t)
\nonumber\\
&&+2\sin{\left({1\over 2}(\partial_r^{\it 1}\partial_k^{\it 2}-
\partial_k^{\it 1}\partial_r^{\it 2})\right)}
\sigma^{\rm hf}(k,r,t)\rho(k,r,t)
\nonumber\\
&&=\exp{\left({i\over 2}\left(\partial_r^{\it 1}\partial_k^{\it 2}-
\partial_k^{\it 1}\partial_r^{\it 2}\right)\right)}
\int\limits_{0}^{\infty}d\tau
\nonumber\\
\times\Bigl(&&
\hat g^>(k,r,t-\frac{\tau}{2},\tau)\hat\sigma^<(k,r,t-\frac{\tau}{2},-\tau)
\nonumber\\
&&+
\hat\sigma^<(k,r,t-\frac{\tau}{2},\tau)\hat g^>(k,r,t-\frac{\tau}{2},-\tau)
\nonumber\\
&&-
\hat g^<(k,r,t-\frac{\tau}{2},\tau)\hat\sigma^>(k,r,t-\frac{\tau}{2},-\tau)
\nonumber\\
&&-
\hat\sigma^>(k,r,t-\frac{\tau}{2},\tau)\hat g^<(k,r,t-\frac{\tau}{2},-\tau)
\Bigr).
\label{timediagonal}
\end{eqnarray}
Here, the superscripts ${\it 1,2}$ denote that the partial derivatives 
apply only to the first and second function in the product, e.g., 
$(\partial_r^{\it 1}\partial_k^{\it 2})^3ab={\partial^3a\over\partial^3r}
{\partial^3b\over\partial^3k}$. 

\subsubsection{Gradient approximation}
The power expansion of the goniometric functions in (\ref{timediagonal}) 
defines the expansion in space gradients. This expansion goes to 
infinite order, but in the following 
we will restrict our attention to the linear 
expansion. It is customary to abbreviate the linear gradient terms with 
the help of the Poisson brackets,
\begin{equation}
2\sin{\left({1\over 2}(\partial_r^{\it 1}\partial_k^{\it 2}-
\partial_k^{\it 1}\partial_r^{\it 2})\right)}ab\approx
\{a,b\}=
{\partial a\over\partial k}{\partial b\over\partial r}-
{\partial a\over\partial r}{\partial b\over\partial k}.
\label{be17}
\end{equation}

The linear approximation in space gradients is straightforward, however,
one has to be careful about time arguments of functions in the collision 
integral of (\ref{timediagonal}). For example, the first and second 
product seem to form an anticommutator in which the linear gradients in 
space cancel. This is not true because of time arguments.

The gradient approximation in time requires a special treatment because 
of the lower integration limit. Due to this limit the integral does not 
define a standard matrix product with respect to the time variables. We
will give this expansion an explicit notation.

In equilibrium, the functions $\hat g^\gtrless$ and $\hat \sigma^\gtrless$ 
do not depend on the center-of-mass time. For slowly evolving systems the
center-of-mass dependence is smooth and weak. Accordingly, in the collision
integral of (\ref{timediagonal}), we can expand the center-of-mass 
time dependence around the time $t$ in powers of $\tau$. Since all functions
in (\ref{timediagonal}) have the same center-of-mass time, $t-{\tau\over 2}$, it is possible 
to write the linear expansion of (\ref{timediagonal}) with respect to time gradients in 
a compact form
\be
&&{\partial \over \partial t}\rho+
\{\epsilon^{\rm hf},\rho\}=I+{\partial \over \partial t}R,
\label{ki}
\ee
with $\epsilon^{\rm hf}={k^2\over 2m}+\sigma^{\rm hf}$. The zero order 
gradient term, $I=I^>-I^<$, reads 
\begin{eqnarray}
I^<&=&\left(1+{i\over 2}\left(\partial_r^{\it 1}\partial_k^{\it 2}-
\partial_k^{\it 1}\partial_r^{\it 2}\right)\right)
\int\limits_0^{\infty}d\tau
\nonumber\\
&\times&\left(\hat\sigma^>(t,\tau)
\hat g^<(t,-\tau)+\hat g^<(t,\tau)\hat\sigma^>(t,-\tau)\right).
\label{i0s}
\end{eqnarray}
$I^>$ is obtained from $I^<$ via the interchange of particles and holes, 
$>\leftrightarrow <$. In the first order gradient term, $R=R^>-R^<$, is
given by
\begin{eqnarray}
R^<&=&-{1\over 2}\left(1+{i\over 2}\left(\partial_r^{\it 1}
\partial_k^{\it 2}-\partial_k^{\it 1}\partial_r^{\it 2}\right)\right)
\int\limits_0^{\infty}\!d\tau\,\tau
\nonumber\\
&\times&\left(\hat\sigma^>(t,\tau)\hat g^<(t,-\tau)+
\hat g^<(t,\tau)\hat\sigma^>(t,-\tau)\right).
\label{icorrs}
\end{eqnarray}

Now we are ready to turn all functions into the Wigner representation 
for time and energy,
\begin{equation} 
\sigma_\omega\left(k,r,t\right)=
\int d\tau\,{\rm e}^{i\omega\tau}\,\hat\sigma(k,r,t,\tau).
\label{wrf}
\end{equation}
We will write the energy argument as the subscript and keep other
arguments implicit in most of the formulas.

After a substitution of $\sigma$'s and $g$'s of the Wigner representation
(\ref{wrf}) into (\ref{i0s}), one can integrate out the difference time 
$\tau$. The integration over time results in the $\delta$ function 
which describes processes on the energy shell, and the principle value 
terms which represent off-shell processes,
\begin{equation}
\int\limits_0^{\infty}\!d\tau\,{\rm e}^{i(\omega-\omega')\tau}=
\pi\delta(\omega-\omega')+i{\wp\over\omega-\omega'}.
\label{wpdef}
\end{equation}
The $\delta$-function can be readily integrated out leaving both functions
with identical energy arguments. The principle value part,$\wp$, is an odd 
function of energies $\omega$ and $\omega'$, therefore its non-gradient 
contributions cancel but the linear gradients survive. Formula (\ref{i0s}) 
thus turns into 
\begin{equation}
I^<=\int{d\omega\over 2\pi}\sigma^>_\omega g^<_\omega + 
\int {d\omega d\omega'\over (2 \pi)^2}{\wp\over\omega'-\omega} 
\left\{\sigma^>_{\omega'},g^<_\omega\right\}.
\label{i0om}
\end{equation}

In the correlated part we express the time integral as
\begin{eqnarray}
\int\limits_0^{\infty}\!d\tau\,\tau\,{\rm e}^{i(\omega-\omega')\tau}&=&
-i{\partial\over\partial\omega}
\int\limits_0^{\infty}\!d\tau\,{\rm e}^{i(\omega-\omega')\tau}
\nonumber\\
&=&-i\pi\delta'(\omega-\omega')+{\wp'\over\omega-\omega'}.
\label{i0omd}
\end{eqnarray}
The meaning of $\delta'$ and $\wp'$ is seen from comparison with (\ref{wpdef}).
The $\delta'$ and $\wp'$ are odd and even functions in $\omega-\omega'$,
respectively. The gradient expansion of (\ref{icorrs}) thus reads
\begin{equation}
R^<=-\int {d\omega d\omega'\over(2\pi)^2}{\wp'\over\omega'-
\omega}\sigma^>_{\omega'}g^<_\omega+\int{d\omega\over 2\pi}
\left\{\sigma^>_\omega,\partial_\omega  g^<_\omega\right\}.
\label{icorrom}
\end{equation}

The term $R$ contributes to the kinetic equation (\ref{ki}) as a gradient 
correction linear in time. Since the second term of $R^<$ in 
(\ref{icorrom}) is proportional to space gradients, we neglect this term 
leading to the second order contribution in gradients.

\subsection{Connection between the Wigner and quasiparticle distributions}
At very low temperatures, dissipative processes are known to vanish due to 
the Pauli exclusion principle. It is desirable to reorganize the kinetic
equation so that the scattering integral will vanish in this limit too.
As can be seen from (\ref{icorrom}), the term $\partial_t R$ remains 
finite even for very low temperatures. It is possible to formally remove 
$\partial_t R$ if we shift this term on the drift side of (\ref{ki}) and 
introduce a new distribution function,
\begin{equation}
f=\rho-R.
\label{redefdist}
\end{equation}
We have denoted the new function as $f$ to anticipate that it is
indeed the 
quasiparticle distribution as will be shown now.

Since we want to arrive at the kinetic equation for $f$, it is advantageous
to write relation (\ref{redefdist}) in the opposite way so that we obtain
$\rho$ as a functional of $f$, $\rho=f+R$. Using (\ref{icorrom}) and the 
particle-hole conjugated term, $R^>$, one finds
\begin{equation}
\rho=f-\int {d\omega'd\omega\over(2\pi)^2}{\wp'\over\omega'-\omega}
\left(\sigma^>_{\omega'}g^<_\omega-\sigma^<_{\omega'}g^>_\omega\right).
\label{an1}
\end{equation}
This relation provides the so called extended quasiparticle approximation
of the Wigner distribution.

\subsubsection{Wave-function renormalization}
To separate the on-shell and off-shell contributions of the correlation
term, we use the imaginary part of the selfenergy, $\gamma=\sigma^>+
\sigma^<$ and the spectral function, $a=g^>+g^<$ to rewrite (\ref{an1}) 
as
\begin{equation}
\rho=f-\int{d\omega'd\omega\over(2\pi)^2}{\wp'\over\omega'-\omega}
\left(\gamma_{\omega'}g^<_\omega-\sigma^<_{\omega'}a_\omega\right).
\label{an1r1}
\end{equation}
With the help of the Kramers-Kronig relation
\begin{equation}
\int{d\omega'\over 2\pi}{\wp'\over\omega-\omega'}\gamma_{\omega'}=
{\partial\sigma_\omega\over\partial\omega},
\label{kk1}
\end{equation}
where $\sigma$ is the real part of the selfenergy, we obtain
\begin{equation}
\rho=f+\int{d\omega\over 2\pi}{\partial\sigma_\omega\over\partial\omega}
g^<_\omega+\int{d\omega'd\omega\over(2\pi)^2}{\wp'\over\omega'-\omega}
\sigma^<_{\omega'}a_\omega.
\label{an1r2}
\end{equation}
The simplest quasiparticle approximation, $a_\omega=2\pi\delta(\omega-
\varepsilon)$ and $g^<_\omega=f\,2\pi\delta(\omega-\varepsilon)$, in the 
correlation terms yields
\begin{equation}
\rho=\left(1+\left.{\partial\sigma\over\partial\omega}\right|_{\omega=
\varepsilon}\right)f+\int{d\omega'\over 2\pi}{\wp'\over \omega'-\varepsilon
}\sigma^<_{\omega'}.
\label{an1r3}
\end{equation}
In (\ref{an1r3}) the on-shell (quasiparticle) contribution is reduced by 
the wave-function renormalization,
\begin{equation}
z=1+\left.{\partial\sigma\over\partial\omega}\right|_{\omega=
\varepsilon},
\label{wfr}
\end{equation}
and the term with $\sigma^<$ provides the off-shell contributions. 

The exact renormalization is $z^{-1}=1-\partial_\omega\sigma$, therefore
(\ref{wfr}) is only the linear approximation in $\partial_\omega\sigma$. 
From the Kramers-Kronig relation (\ref{kk1}) one can see that this
corresponds to the linear approximation in the scattering rate $\gamma$.
We will keep all terms only up to this order in $\gamma$.

\subsubsection{Extended quasiparticle approximation}
From the relation between the Wigner and the quasiparticle distributions 
(\ref{an1r3}) one can deduce an approximative construction of the 
correlation function $g^<$ as a functional of the quasiparticle 
distribution $f$. Since $\rho=\int{d\omega\over 2\pi} g^<$, we can write 
(\ref{an1r3}) as
\begin{equation}
\int{d\omega\over 2\pi}g^<_\omega=\int{d\omega\over 2\pi}\Biggl(
f\,z\,2\pi\delta(\omega-\varepsilon)+{\wp'\over \omega-\varepsilon}
\sigma^<_\omega\Biggr).
\label{an1r4}
\end{equation}
We have multiplied $f$ with the $\delta$ function and included this term
into the integrand. 

One can see that the correlation function in the extended quasiparticle
approximation \cite{SL94,SL95},
\begin{equation}
g^<_\omega=f\,z\,2\pi\delta(\omega-\varepsilon)+{\wp'\over\omega-
\varepsilon}\sigma^<_\omega,
\label{ansatz2}
\end{equation}
satisfies (\ref{an1r4}). Perhaps it is not necessary to recall that 
the function $f$ has been introduced here merely to suppress the
non-dissipative part of the collision integral, $\partial_tR$. In 
\cite{SL94,SL95}, the function $f$ has been defined as the factor of the
singularity of $g^<$ which justifies its interpretation in terms of the 
quasiparticle distribution. As relation (\ref{an1r4}) shows, both 
approaches lead to identical results.

In parallel with (\ref{ansatz2}), for the hole correlation function one 
finds
\begin{equation}
g^>_\omega=(1-f)\,z\,2\pi\delta(\omega-\varepsilon)+{\wp'\over\omega-
\varepsilon}\sigma^>_\omega.
\label{ansatz2h}
\end{equation}
From the spectral identity, $a=g^>+g^<$, it follows that (\ref{ansatz2}) 
and (\ref{ansatz2h}) are consistent with the extended quasiparticle 
approximation of the spectral function 
\begin{equation}
a_\omega=z\,2\pi\delta(\omega-\varepsilon)+{\wp'\over\omega-
\varepsilon}\gamma_\omega.
\label{ansatz2s}
\end{equation}

It is worthy to remark that (\ref{ansatz2s}) fulfills the spectral sum 
rule \cite{KKL84}
\beq\label{suma}
\int \frac{d\omega}{2\pi} a(k,\omega,r,t) &=&1
\eeq
as well as the energy weighted sum rule \cite{LSM97}
\beq\label{sumb}
\int \frac{d\omega}{2\pi} \omega a(k,\omega,r,t) 
&=&{k^2 \over 2 m}+\sigma^{\rm hf}(k,r,t).
\eeq
The latter one is violated in the simple quasiparticle picture.

\subsection{Landau-Silin equation}
So far we have treated only the time gradients finding that the identity
(\ref{ki}) can be expressed as
\begin{equation}
{\partial f\over\partial t}+\{\epsilon^{\rm hf},\rho\}=I.
\label{kimix}
\end{equation}
Now we rearrange this identity into the Landau-Silin equation for $f$.

The time-local remainder $I$ of the collision integral still includes the
space gradients, $I=I_B-I_\nabla$. The non-gradient part [the first term 
of (\ref{i0om})] is the Boltzmann-type scattering integral of the 
Landau-Silin equation
\beq
I_B&=&\int{d\omega\over 2\pi}
(g^>_\omega\sigma^<_\omega-g^<_\omega\sigma^>_\omega)
=z\left((1-f)\sigma^<_\varepsilon-f\sigma^>_\varepsilon\right)
\nonumber\\
\label{o}
\eeq
In the second step we have used (\ref{ansatz2}). The off-shell  
contributions to (\ref{o}) have not been neglected but cancel exactly.

The gradient part which is the second term of (\ref{i0om}) reads
\be
I_\nabla&=&-\int{d\omega d\omega'\over(2\pi)^2}
{\wp\over\omega'-\omega}\left(
\left\{\sigma^>_{\omega'},g^<_\omega\right\}-
\left\{\sigma^<_{\omega'},g^>_\omega\right\}\right)\nonumber\\
&=&\int{d\omega\over(2\pi)}
\left(
\left\{\sigma_{\omega},g^<_\omega\right\}-
\left\{\sigma^<_{\omega},g_\omega\right\}\right).
\label{i0omr1}
\ee
Here (\ref{kk1}) has been used to remove the $\omega'$-integration for
$\sigma^>+\sigma^<$ and a similar Kramers-Kronig relation for the
$\omega$-integration over the spectral function $g^>+g^<$ has been
applied. 
The function
$g={\rm Re}g^{R,A}={1\over 2}(g^R+g^A)$ is the off-shell part of the 
propagators.

The mean-field drift $\{\epsilon^{\rm hf},\rho\}$ in (\ref{kimix}) is not 
compatible with the quasiparticle distribution in the time derivative and
the scattering integral. Moreover, the space gradients from the collision 
integral, $I_\nabla$, have to be accounted for. Our aim is now to show 
that all gradients can be collected into the quasiparticle drift,
\begin{equation}
\{\epsilon^{\rm hf},\rho\}+I_\nabla=\{\varepsilon,f\}.
\label{gradorg}
\end{equation}
The quasiparticle energy differs from the mean-field energy by the pole 
value of the selfenergy, $\varepsilon=\epsilon^{\rm hf}+\sigma_\varepsilon$.

We will proceed in two steps. First we observe that the right hand side of
(\ref{gradorg}) includes only the on-shell contribution, therefore we show
that the on-shell part of the left hand side equals the right hand side.
Second, we show that the off-shell part of the left hand side of (\ref{gradorg})
vanishes. The on-shell parts of (\ref{ansatz2}) used in 
(\ref{i0omr1}) result into
\be
I_\nabla^{\rm on}&=&\int d\omega \{\sigma_\omega,z f
\delta(\omega-\varepsilon)\}\nonumber\\
&=&\{\sigma_\varepsilon,zf\}-\sigma_\varepsilon'\{\varepsilon,zf\}+z
f \{\sigma_\varepsilon',\varepsilon\}\nonumber\\
&=&\{\sigma_\varepsilon,zf\}+(1-z)\{\varepsilon,zf\}+z
f \{z-1,\varepsilon\}.
\ee
Now we add the on-shell part of the commutator $\{\epsilon^{\rm hf},\rho
\}^{\rm on}=\{\epsilon^{\rm hf},zf\}$. Using a rearrangement
\be
&&\{\epsilon^{\rm hf},z
f\}=\{\varepsilon-\sigma,zf\}\nonumber\\
&&=\{\varepsilon,f\}+(z-1)\{\varepsilon,f\}+f
\{\varepsilon,z\}-\{\sigma,z f\}
\ee
we obtain 
\be
\{\epsilon^{\rm hf},\rho\}^{\rm on}+I_\nabla^{\rm on}
=\{\varepsilon,f\}-\{\varepsilon,(1-z)^2 f\}.
\label{relat2}
\ee
The last term is of higher order than linear in the damping and can be
neglected which confirms the relation (\ref{gradorg}) already from the
on-shell parts. 

It remains to prove that all off-shell contributions to the left hand side
of (\ref{gradorg}), ${\cal O}=\{\epsilon^{\rm hf},\rho\}^{\rm off}+
I_\nabla^{\rm off}$ cancel, i.e., ${\cal O}=0$. From the off-shell term 
of (\ref{ansatz2}) we find
\begin{equation}
{\cal O}=\int{d\omega\over 2\pi}\Biggl(\left\{\epsilon^{\rm hf}+
\sigma_\omega,{\wp'\over\omega-\varepsilon}\sigma^<_\omega\right\}-
\left\{g_\omega,\sigma^<_\omega\right\}\Biggr).
\label{i0omr3}
\end{equation}
The term with $\epsilon^{\rm hf}$ results from $\{\epsilon^{\rm 
hf},\rho\}$, while the others from $I_\nabla$. In the last term we use
the extended quasiparticle approximation of the real part of the propagator
\be
g_{\omega}=z {{\wp}\over \omega-\varepsilon}+{{\wp'}\over
\omega-\varepsilon}(\sigma_\omega-\sigma_\varepsilon)
\label{h1}
\ee
which directly results from (\ref{ansatz2s}) and the Kramers-Kronig 
relation. Using 
$\left\{{\wp \over \omega-\varepsilon},\sigma^<_\omega\right\}=-
\left\{\varepsilon,{\wp'\over\omega-\varepsilon}\sigma^<_\omega\right\}$
we can rewrite ${\cal O}$ as
\be
&&{\cal O}=\nonumber\\
&&\int{d\omega\over 2\pi}\Biggl(\left\{\sigma_\omega-\sigma_\varepsilon,
{\wp'\over\omega-\varepsilon}\sigma^<_\omega\right\}
-\left\{{\wp'\over\omega-\varepsilon}(\sigma_\omega-\sigma_\varepsilon),
\sigma^<_\omega\right\}
\Biggr).\nonumber\\
&&
\label{i0omr4}
\ee
The linear expansion in the vicinity of the pole, 
$\sigma_\omega-\sigma_\varepsilon=(\omega-\varepsilon)(z-1)+o(\gamma^2)$,
yields
\be
{\cal O}=\int{d\omega\over 2\pi}\left\{{\wp'' \over
  \omega-\varepsilon},(z-1) \sigma_\omega^<\right\}.
\label{i0omr5}
\ee
The product $(z-1)\sigma^<$ is of second order in $\gamma$, i.e., the 
off-shell contribution ${\cal O}$ is negligible within the assumed 
accuracy. This completes the proof of relation (\ref{gradorg}).

In summary, the requirement that the scattering integral vanishes at very 
low temperatures directly leads to the concept of quasiparticles 
represented by relation (\ref{an1}) between the Wigner and the
quasiparticle distributions
\begin{equation}
\rho=f+\int{d\omega\over 2\pi} {\wp\over\omega-\varepsilon}
{\partial\over\partial\omega}
\left((1-f)\sigma^<_\omega-f\sigma^>_\omega\right).
\label{tr2}
\end{equation}
The space gradients of the collision integral renormalize the mean-field 
drift into the familiar quasiparticle drift (\ref{gradorg}). The resulting 
kinetic equation for the quasiparticle distribution has the structure of the
Landau-Silin equation (\ref{tr1})
\begin{equation}
{\partial f\over\partial t}+{\partial\varepsilon\over\partial k}
{\partial f\over\partial r}-{\partial\varepsilon\over\partial r}
{\partial f\over\partial k}=z((1-f)\sigma^<_\varepsilon -
f\sigma^>_\varepsilon).
\label{Silin}
\end{equation}

We conclude this section that the Levinson type of equation (\ref{levinson})
is equivalent to the Landau-Silin equation (\ref{tr1}) up to second
order in the damping $\gamma$ or in the extended quasiparticle picture.

\section{Additional retardation of the Levinson equation}
The link between the Levinson and Landau-Silin equations shows that
the collision integral of the Levinson equation cannot be fully 
interpreted in terms of the scattering processes but it includes 
various renormalization features. In the above treatment we
have constructed a closed equation for the quasiparticle distribution
function while the reduced density is given by an additional
functional of this quasiparticle distribution. There arises an 
important question whether it is not possible to construct the 
correlation functions, $g^{>,<}$, directly from the Wigner 
distribution so that identity (\ref{timediagonal}) turns into 
a closed equation for the Wigner distribution. To show the problems
and drawbacks with this opposite way we discuss this 
construction for a homogeneous system.

An approximation giving the Levinson equation is the GKB ansatz
\cite{LSV86} which for the homogeneous system reads [$\tau>0$]
\begin{eqnarray}
\hat g^<(k,t-{\tau\over 2},\tau)&=&i\hat g^R(k,t-{\tau\over 2},\tau)
\rho(k,t-\tau),
\nonumber\\
\hat g^<(k,t-{\tau\over 2},-\tau)&=&-i\hat g^A(k,t-{\tau\over 2},-\tau)
\rho(k,t-\tau).
\label{gkban}
\end{eqnarray}
For $\hat g^>$ one substitutes $1-\rho$ for $\rho$. Note that
this approximation includes the additional retardation between the
center-of-mass time $t-{\tau\over 2}$ and the time argument of the Wigner
distribution which is $t-\tau$. This additional retardation causes many
problems ranging from instability of numerical treatments over double 
counts of renormalizations or virial corrections to incorrect 
interpretations of various correlation phenomena. Here we show how to
handle this retardation within the linear approximation.

If one assumes that the propagator $G^R$ depends only on the difference 
time $\tau$, i.e., it does not depend on any time dependent external 
fields or mean fields, it is possible to follow the approach developed
above with only minor modifications. To make our discussion specific we 
limit our attention to homogeneous systems and employ the Galitskii-Feynman
T-matrix approximation of the selfenergy,
\begin{eqnarray}
&&\hat\sigma^<(k,t-{\tau\over 2},\tau)
\nonumber\\
&=&\int{dpdq\over(2\pi)^6}\hat g^>(p,t-{\tau\over 2},-\tau)
\nonumber\\
&\times&\int d\tau_R d\tau_A
{\cal T}^R(k,p,q,\tau_R){\cal T}^A(k,p,q,-\tau_A)
\nonumber\\
&\times&\hat g^<\left(k-q,t-{1\over 2}(\tau_R+\tau_A+\tau),
\tau-\tau_R+ \tau_A\right)
\nonumber\\
&\times&\hat g^<\left(p+q,t-{1\over 2}(\tau_R+\tau_A+\tau),
\tau-\tau_R+ \tau_A\right).
\label{le10}
\end{eqnarray}
This selfenergy describes the process in which two particles of initial
momenta $k-q$ and $p+q$ are scattered into final states $k$ and $p$.
The function ${\cal T}^{R,A}$ is the retarded/advanced scattering
T-matrix, see appendix~A. 
Due to a finite duration of the scattering process, the T-matrices 
bring yet additional retardation times $\tau_{R,A}$. We will show that
retardation times $\tau_{R,A}$ are responsible for the virial corrections
to the density.

\subsection{Decay of propagators and double counts}
First we focus on double counts following from the additional retardation 
resulting from the GKB ansatz. Since these problems appear already for
weakly coupled systems, it is sufficient to restrict our attention to the
Born approximation, ${\cal T}^{R,A}={\cal V}\delta(\tau_{R,A})$. The
Levinson equation then reads
\begin{equation}
{\pa t}\rho(k,t)=2\,{\rm Re}\int{dpdq\over(2\pi)^6}{\cal V}_q^2
\int\limits_{0}^{\infty}d\tau S(t-{\tau\over 2},\tau)D(t-\tau).
\label{levred}
\end{equation}
All distributions are collected in
\begin{eqnarray}
D(t)&=&(1\!-\!\rho(k,t))(1\!-\!\rho(p,t))\rho(k-q,t)\rho(p+q,t)
\nonumber\\
&-&\rho(k,t)\rho(p,t)(1\!-\!\rho(k-q,t))(1\!-\!\rho(p+q,t)),
\label{timediagonal2}
\end{eqnarray}
all propagators are collected in
\begin{eqnarray}
S(t,\tau)&=&\hat g^R(k,t,\tau)\hat g^R(p,t,\tau)
\nonumber\\
&\times&\hat g^A(k-q,t,-\tau)\hat g^A(p+q,t,-\tau).
\label{props}
\end{eqnarray}
We have used that the Wigner distribution is a real function and the 
complex conjugacy of propagators, $\hat g^R(k,t,\tau)=\left(\hat 
g^A(k,t,-\tau)\right)^*$, to reduce the expressions.

Let us assume for the moment that the propagators do not depend on the 
center-of-mass time $t$. In this case the only $t$ dependence in the 
collision integral of the Levinson equation (\ref{levred}) is due to
distributions. Since for the Born approximation all distributions are 
retarded in an equal way, we can follow a modification of the approach 
used above. 

For slow perturbations one can expand (\ref{levred}) to the lowest 
order in the memory, $D(t-\tau)=D(t)-\tau\partial_tD(t)$, so that the
collision integral splits into the Boltzmann-like scattering integral
and the gradient correction, $\partial_t\rho={\cal I}+\partial_t{\cal R}$, 
with
\begin{eqnarray}
{\cal I}&=&\int{dpdq\over(2\pi)^6}{\cal V}_q^2D(t)
2\,{\rm Re}\int\limits_{0}^{\infty}d\tau S(\tau),
\label{lowordle}\\
{\cal R}&=&\int{dpdq\over(2\pi)^6}{\cal V}_q^2D(t)
2\,{\rm Re}\int\limits_{0}^{\infty}d\tau\,\tau\,S(\tau).
\label{firordle}
\end{eqnarray}
Please note that ${\cal R}$ and ${\cal I}$ are different from
(\ref{ki}). Again, from (\ref{levred}) we can define a new function to remove 
the retardation,
\begin{equation}
{\cal F}=\rho-{\cal R},
\label{newf}
\end{equation}
and try to redefine the transport in terms of ${\cal F}$. This will us
lead now to unavoidable double counts.

\subsubsection{Double counts}
The integrals in (\ref{lowordle}) are controlled by two mechanisms, the
de-phasing given by momentum integrals and the decay of propagators. Let
us assume for a while that the de-phasing is the dominant part leaving the 
decay aside. Within the quasiparticle approximation, $\hat g^R(k,\tau)
\approx-i{\rm e}^{-i\varepsilon_k\tau}$, which is free of the decay, the 
time integrals over propagators yield 
\begin{eqnarray}
2\,{\rm Re}\int\limits_{0}^{\infty}d\tau S_{\rm qp}(\tau)&=&2\pi\delta
\left(\varepsilon_k+\varepsilon_p-\varepsilon_{k-q}-\varepsilon_{p+q}\right),
\label{le12a}\\
2\,{\rm Re}\int\limits_{0}^{\infty}d\tau\,\tau\, S_{\rm qp}(\tau)&=&-
2{\wp'\over\varepsilon_k+\varepsilon_p-\varepsilon_{k-q}-\varepsilon_{p+q}}.
\label{le12b}
\end{eqnarray}
After approximation (\ref{le12a}), ${\cal I}$ from (\ref{lowordle}) turns 
into the Boltzmann-type scattering integral $I_B$ of the Landau-Silin 
equation. Naturally, the on-shell contributions are identical since 
they do not depend on the retardation.

Using (\ref{le12b}) in (\ref{firordle}) one finds 
\begin{equation}
{\cal R}=2\int{dpdq\over(2\pi)^6}{\cal V}_q^2D(t)
{\wp'\over\varepsilon_k+\varepsilon_p-\varepsilon_{k-q}-\varepsilon_{p+q}}.
\label{nfrho1}
\end{equation}
The function ${\cal R}$ is twice the off-shell contribution $R$ to the
Wigner distribution. To show this, we express (\ref{nfrho1}) in terms 
of the Born selfenergy (\ref{le10}) and the approximative correlation 
functions, $g^<=2\pi\delta(\omega-\varepsilon)\rho$, as
\begin{equation}
{\cal R}=-2\int {d\omega'd\omega\over(2\pi)^2}{\wp'\over\omega-\omega'}
\left(\sigma^>_{\omega'}g^<_\omega-\sigma^<_{\omega'}g^>_\omega\right).
\label{nan1}
\end{equation}
Comparing with the second term of (\ref{an1}) one can see that the 
additional retardation results in the double count of the correlated part,
${\cal R}=2R$. Due to this double count, the function ${\cal F}=\rho-
{\cal R}=\rho-2R=f-R$ cannot be interpreted as the distribution of 
excitations because it has large regions of negative values. 
Please note a common met pitfall in literature related to this double
count given in appendix~C.

We note that in numerical treatments of the Levinson equation 
neglecting or underestimating the decay, leads to a numerical 
instability of the equation, as it has been reported by Haug \cite{BTRH92}.
Other numerical solutions show a continuous increase of kinetic energy, 
e.g. figure~4 of \cite{KBKS97} or pulsation modes \cite{P98}. 
The instability of the solution is a problem which shows that the Levinson 
equation is a difficult tool to handle.
The double count of the off-shell contribution resulting in the negative
effective occupation factors is likely one of the reasons of this problem.
A detailed discussion and numerical comparison of the Levinson
equation with the Kadanoff and Baym equation can be found in \cite{KM01}.

\subsubsection{Decay of propagators}
The double count of correlations is a serious mistake for theories which
aim to go beyond the Boltzmann equation. Here we show that the decay of 
propagators during the collision removes this double count.

We use the simplest approximation of the decay, 
\begin{equation}
\hat g^R\left(k,t,\tau\right)=-i{\rm e}^{-i\varepsilon_k\tau}
{\rm e}^{-\gamma_k{\tau\over 2}},
\label{proppol}
\end{equation}
where $\varepsilon_k=\varepsilon(k,t-{\tau\over 2})$ and $\gamma_k=\gamma_{
\varepsilon(k,t-{\tau\over 2})}(k,t-{\tau\over 2})$ is the pole value of 
$\gamma=\sigma^>+\sigma^<=-2{\rm Im}\,\sigma^R$. The time argument shows 
that external or mean fields can modify the propagation during the 
collision. The collected propagator thus reads
\begin{equation}
S\left(t-{\tau\over 2},\tau\right)=S_{\rm qp}\left(t-{\tau\over 2},\tau\right)
{\rm e}^{-(\gamma_k+\gamma_p+\gamma_{k-q}+\gamma_{p+q}){\tau\over 2}}.
\label{colprop}
\end{equation}

The kinetic equation holds only if the collision duration is short on the
scale of the lifetime, $(2\gamma)^{-1}$. In this limit we can expand the
exponential decay,
\begin{equation}
{\rm e}^{-(\gamma_k+\gamma_p+\gamma_{k-q}+\gamma_{p+q}){\tau\over 2}}=
1-(\gamma_k+\gamma_p+\gamma_{k-q}+\gamma_{p+q}){\tau\over 2}.
\label{colpropex}
\end{equation}
Since we are dealing with the correction terms, we can use a simple
approximation of the kinetic equation, $\partial_t\rho_k=\sigma^<_k-
\gamma_k\rho$, to rearrange the contribution of $\gamma$'s to its 
integrand, 
\begin{equation}
-{\tau\over 2}(\gamma_k+\gamma_p+\gamma_{k-q}+\gamma_{p+q})=
{\tau\over 2}{\partial\over\partial t}D-{\tau\over 2}D_{\rm 3p}.
\label{gamtoder}
\end{equation}
The last term, 
\begin{eqnarray}
&&D_{\rm 3p}=\sigma^>_k(1\!-\!\rho_p)\rho_{k-q}\rho_{p+q}+
(1\!-\rho_k)\sigma^>_p\rho_{k-q}\rho_{p+q}
\nonumber\\
&&+(1\!-\!\rho_k)(1\!-\!\rho_p)\sigma^<_{k-q}\rho_{p+q}+
(1\!-\!\rho_k)(1\!-\!\rho_p)\rho_{k-q}\sigma^<_{p+q}
\nonumber\\
&&-\sigma^<_k\rho_p(1\!-\!\rho_{k-q})(1\!-\!\rho_{p+q})-
\rho_k\sigma^<_p(1\!-\!\rho_{k-q})(1\!-\!\rho_{p+q})
\nonumber\\
&&-\rho_k\rho_p\sigma^>_{k-q}(1\!-\!\rho_{p+q})-
\rho_k\rho_p(1\!-\!\rho_{k-q})\sigma^>_{p+q},
\label{3pdef}
\end{eqnarray}
can be neglected leading to the three-particle contribution to the 
scattering integral. Please note that such neglects have not been
necessary in our suggested construction $\rho[f]$ outlined in Sec.~II. 

Within the linear approximation in time gradients one finds that the
decay reduces the retardation,
\begin{equation}
{\rm e}^{-(\gamma_k+\gamma_p+\gamma_{k-q}+\gamma_{p+q}){\tau\over 2}}
D(t-\tau)=D\left(t-{\tau\over 2},\tau\right).
\label{redfin}
\end{equation}
Now the collective propagator and distribution have the same time
retardation of the center-of-mass time being equal to ${\tau\over 2}$.
This time shift removes the double count and reproduces the result 
obtained already in Sec.~II.

We note that the original Levinson equation \cite{L65} does not include 
the decay of propagators in the scattering integral. The presented discussion
shows that the decay plays the important role for the correct treatment 
of renormalizations and other contributions of correlations.

\subsection{Collision delay}
We have seen that within the Born approximation the retarded
collision integral of the Levinson equation can be transformed into the
instantaneous scattering integral of the Boltzmann type and quasiparticle
renormalizations of the single-particle distribution and energy. Beyond
the Born approximation, there is the additional retardation expressed by
times $\tau_{R,A}$ in the selfenergy (\ref{le10}). This retardation is 
responsible for the virial correction, in particular for the correlated
density $n_c$. Since the retardation times $\tau_{R,A}$ do not enter 
the final states of the collision, they cannot be removed in the above
manner. On the other hand, the time integrals over $\tau_{R,A}$ can be 
eliminated so that the collision duration is expressed by an effective 
time-non-locality of the scattering integral. For simplicity of presentation
in this section we assume that T-matrices and selfenergies are independent 
of the center-of-mass time. This will lead us to the basic idea. The
complete result with all time and space dependences are given in the
next Sec.~IV.

Within the linear approximation in time gradients we can treat the 
contributions of times $\tau_{R,A}$ separately from other retardations.
We first implement the quasiparticle ansatz and expand in retardations,
\begin{eqnarray}
&&\hat g^<\left(k-q,t-{1\over 2}(\tau_R+\tau_A+\tau),
\tau-\tau_R+ \tau_A\right)
\nonumber\\
&\times&\hat g^<\left(p+q,t-{1\over 2}(\tau_R+\tau_A+\tau),
\tau-\tau_R+ \tau_A\right)
\nonumber\\ 
&=&\left(1-{1\over 2}\tau{\partial\over\partial t}-{1\over 2}
(\tau_R+\tau_A){\partial\over\partial t}\right)
\nonumber\\ 
&\times&f_{k-q}(t)f_{p+q}(t)
{\rm e}^{-i(\varepsilon_{k-q}+\varepsilon_{p+q})(\tau-\tau_R+\tau_A)}.
\label{sepgr}
\end{eqnarray}
The contribution of ${1\over 2}\tau{\partial\over\partial t}$ has been
discussed above, now we focus on the term due to ${1\over 2}
(\tau_R+\tau_A){\partial\over\partial t}$.

Introducing the energy representation of the T-matrix,
\begin{equation}
{\cal T}^R(\tau_R)=\int{d\omega_R\over 2\pi}t^R(\omega_R)
{\rm e}^{-i\omega\tau_R},
\label{WT}
\end{equation}
and a complex conjugate for ${\cal T}^A$, the contribution of ${1\over 2}
(\tau_R+\tau_A){\partial\over\partial t}$ to the selfenergy (\ref{le10}) reads
\begin{eqnarray}
\Delta\hat\sigma^<(k,t,\tau)
&=&-{1\over 2}\int{dpdq\over(2\pi)^6}{\rm e}^{i\varepsilon_p\tau}(1-f_p)
\nonumber\\
&\times&
\int{d\omega_Rd\omega_A\over(2\pi)^2}
t^R(k,p,q,\omega_R)t^A(k,p,q,\omega_A)
\nonumber\\
&\times&\int d\tau_R d\tau_A
{\rm e}^{-i\omega_R\tau_R}{\rm e}^{i\omega_A\tau_A}(\tau_R+\tau_A)
\nonumber\\ 
&\times&{\partial\over\partial t}f_{k-q}f_{p+q}
{\rm e}^{-i(\varepsilon_{k-q}+\varepsilon_{p+q})(\tau-\tau_R+\tau_A)}.
\label{le10D}
\end{eqnarray}
The integral over $\tau_{R,A}$ results in the derived $\delta$-functions
which can be readily integrated out,
\begin{eqnarray}
\Delta\hat\sigma^<(k,t,\tau)
&=&\int{dpdq\over(2\pi)^6}
{\rm e}^{i(\varepsilon_p-\varepsilon_{k-q}-\varepsilon_{p+q})\tau}(1-f_p)
\nonumber\\
&\times&{1\over 2i}\left({\partial t^R\over\partial\omega}t^A-t^R
{\partial t^A\over\partial\omega}\right)_{\omega=\varepsilon_{k-q}
+\varepsilon_{p+q}}
\nonumber\\ &\times&
{\partial\over\partial t}f_{k-q}f_{p+q}.
\label{le10Di}
\end{eqnarray}

The correction (\ref{le10Di}) is proportional to the collision delay defined 
from the phase shift,
\begin{equation}
\Delta_t={\partial\phi\over\partial\omega}.
\label{Wtime}
\end{equation}
Indeed, writing the T-matrices in the form $t^{R,A}=|t|{\rm e}^{\mp i\phi}$, 
where $|t|^2$ is the amplitude of the scattering rate and $\phi$ is the 
scattering phase shift, we obtain
\begin{equation}
{1\over 2i}\left({\partial t^R\over\partial\omega}t^A-t^R
{\partial t^A\over\partial\omega}\right)=-|t|^2\Delta_t.
\label{Dphi}
\end{equation}
This allows us to join the correction with the time-local part of the
selfenergy,
\begin{eqnarray}
\hat\sigma^<(k,t,\tau)
&=&\int{dpdq\over(2\pi)^6}|t|^2
{\rm e}^{i(\varepsilon_p-\varepsilon_{k-q}-\varepsilon_{p+q})\tau}
\left(1-f_p(t)\right)
\nonumber\\ &\times&
f_{k-q}(t-\Delta_t)f_{p+q}(t-\Delta_t).
\label{le10t}
\end{eqnarray}

The selfenergy (\ref{le10t}) can be compared with (\ref{le10}). First, the
retardation by $-\tau/2$ has been suppressed because it is eliminated by
the method described in Sec.~II. Second, the retardations by the internal 
times of the T-matrices, $\tau_{R,A}$, have been reduced from the integral
form of (\ref{le10}) to the effective time non-locality of (\ref{le10t}).
The latter form includes only a single characteristic time $\Delta_t$
which is similar (but not fully identical) to the collision delay 
defined by Wigner from the scattering phase shift. Note that in form
(\ref{le10t}) no time integration appears. 

An interpretation of the collision delay $\Delta_t$ is troubled by the
fact that this effective time does not have to be positive. Apparently, one
cannot interpret the collision delay as a mean time scale of the $\tau_{R,A}$ 
integrals in the selfenergy (\ref{le10}). The negative value of $\Delta_t$
appears when the phase shift decreases with increasing energy, as it
happens above a bound state or a resonant scattering state. For instance,
the collision delay of nucleon-nucleon collisions is negative at low 
relative energies due to the deuteron bound state, see \cite{MLSK98}. 
The negative value of the collision delay merely tells that particles 
anti-correlate at a given relative energy. 

\section{Nonlocal scattering integral}
The selfenergy (\ref{le10t}) is easily converted into the scattering integral. 
The $|t|^2$ represents the scattering amplitude in the T-matrix approximation. 
The exponential function turns into the energy conserving $\delta$-function. 
Compared to the common scattering integral of the Landau-Silin equation, the 
only difference consists in the collision delay $\Delta_t$ entering the initial 
states of the collision. Of course, in the previous section we have assumed a 
homogeneous system with time-independent selfenergies and T-matrices. For a 
general case more nonlocal corrections appear. These corrections we discuss
in this section.

In this section we thus focus on the scattering integral. Our aim is to 
derive the scattering integral within the same approximations as the skeleton 
kinetic equation discussed in Sec.~II, i.e., keeping all linear gradients. 
This procedure leads to the non-instantaneous and non-local 
scattering integral. We use the Galitskii-Feynman T-matrix approximation 
of the selfenergy which is appropriate for dense systems of particles
interacting via short range potentials. The T-matrix approximation for 
non-equilibrium Green functions is briefly presented in Appendix \ref{at}.

\subsection{Quasi-classical limit}
The consistent treatment of gradient contributions is provided by the
quasi-classical limit, i.e., the linear expansion in gradients. The collision
delay derived above shows that beside gradient contributions forming the
drift in the skeleton equation, there are non-trivial gradient contributions
making the scattering integral non-local in time. Transforming the system 
into a running coordinate framework, one can see that the collision delay 
has be accompanied by the space non-locality of the collision integral. 
Accordingly, the gradient expansion has to be applied also to all internal 
space and time integrations of the scattering integral. 

The quasi-classical limit and the limit of small scattering rates
explicitly determine how to evaluate the scattering integral from the
selfenergy $\sigma^<$. For non-degenerate systems, a very similar scheme 
was carried out by B{\"a}rwinkel \cite{B69,B69a}. One can see in 
B{\"a}rwinkel's papers, that the scattering integral is troubled by a 
large set of gradient corrections. This formal complexity seems to be 
the main reason why most authors either neglect gradient corrections at 
all \cite{bm90,D84} or provide them buried in multi-dimensional integrals 
\cite{L90,L90a,MR95a}. For a degenerate system, the set of gradient 
corrections to the scattering integral is even larger than for rare gases 
studied by B{\"a}rwinkel, see \cite{SLM96}. To avoid manipulations
with long and obscure formulas, the gradient corrections have to be
sorted and expressed in a comprehensive form.  To this end, we will 
express all gradient corrections in the form of shifted arguments, as we 
have done above for the collision delay.

The quasi-classical limit of the scattering integral with all linear 
gradients kept is a tedious but straightforward algebraic exercise, see 
Appendix \ref{expansion}, giving
\begin{eqnarray}
&&\sigma^<(\omega,k,r,t)
\nonumber\\
&&=\int\!{dp\over(2\pi)^3}{dE\over 2\pi}
{dq\over(2\pi)^3}{d\Omega\over 2\pi}
\Biggl(1-{1\over 2}{\partial\Delta_2\over\partial r}\Biggr)
\nonumber\\
&&\times
\Biggl|t\Bigl(\!\omega\!+\!E\!-\!\Delta_E,k\!-\!{1\over 2}\!
\Delta_K,p\!-\!{1\over 2}\!\Delta_K,q,r\!-\!\Delta_r,t\!-\!{1\over 2}\!
\Delta_t\Bigr)\!\Biggr|^2
\nonumber\\
&&\times
g^>(E,p,r-\Delta_2,t)
\nonumber\\
&&\times
g^<(\omega-\Omega-\Delta_E,k-q-\Delta_K,r-\Delta_3,t-\Delta_t)
\nonumber\\
&&\times
g^<(E+\Omega-\Delta_E,p+q-\Delta_K,r-\Delta_4,t-\Delta_t).
\label{vc7d}
\end{eqnarray}
With $g^{>,<}$ substituted from the quasiparticle approximation, this 
formula turns into the functional of the quasiparticle distribution. As
one can see, the initial states, $k-q$ and $p+q$, include the collision
delay $\Delta_t$, as in the homogeneous system describe by (\ref{le10t}). 

Beside the collision delay, non-local selfenergy (\ref{vc7d}) includes
a number of other gradient corrections which enter nearly all arguments.
These are expressed via $\Delta$'s which are derivatives of the scattering 
phase shift $\phi$, where $t^{R,A}=|t|{\rm e}^{\mp i\phi}$, according to 
the following list \cite{SLM96,LSM97}
\begin{eqnarray}
\Delta_2&=&{\partial\phi\over\partial p}-{\partial\phi\over\partial q}-
{\partial\phi\over\partial k},
\label{vc4a}\\
\Delta_3&=&-{\partial\phi\over\partial k},
\label{vc4b}\\
\Delta_4&=&-{\partial\phi\over\partial q}-{\partial\phi\over\partial k},
\label{vc4c}\\
\Delta_t&=&{\partial\phi\over\partial\omega},
\label{vc4d}\\
\Delta_K&=&{1\over 2}{\partial\phi\over\partial r},
\label{vc4e}\\
\Delta_E&=&-{1\over 2}{\partial\phi\over\partial t},
\label{vc4f}\\
\Delta_r&=&{1\over 4}\left(\Delta_2+\Delta_3+\Delta_4\right).
\label{vc4g}
\end{eqnarray}
Finally, there is the energy gain which is discussed in \cite{LSM99}. 
Numerical values of the delay $\Delta_t$ and the displacements $\Delta_{2-4}$
for the scattering of isolated particles are presented in \cite{MLSK98}.

Sending all $\Delta$'s to zero one recovers the instantaneous and local
approximation of the selfenergy, Eq.~(\ref{vc7d}). The non-instantaneous 
and non-local corrections given by $\Delta$'s appear in three ways: in
arguments of the correlation functions, in arguments of the T-matrix, and 
as a fore-factor. In correlation functions, the $\Delta$'s describe 
displacements of the initial and final positions of colliding particles,
the final state of one of the particles is fixed to the balanced phase-space 
point $(k,r,t)$. The arguments of the T-matrix correspond to the center of
arguments of all initial and final states. 

The pre-factor has no special 
physical meaning as it depends on the actual choice of energy and momentum
variables. For instance, using in (\ref{vc7d}) an integral over $E'=E-
\Delta_E$ instead of the integral over $E$, a corresponding factor, 
$dE/dE'=1+\partial_\omega\Delta_E=1-{1\over 2}\partial_t\Delta_t$, appears.

\subsection{Extended quasiparticle approximation}
Now we can complete the derivation of the kinetic equation. Since the
scattering integral is already proportional to the `small' scattering rate, we 
can use from the extended quasiparticle approximation (\ref{ansatz2}) the
pole part, $g^<(\omega,k,r,t)=f(k,r,t)2\pi
\delta(\omega-\varepsilon(k,r,t))$, to convert selfenergies $\sigma^{>,<}$ 
into functionals of the quasiparticle distribution. The selfenergy $\sigma^<$ 
is given by (\ref{vc7d}) and $\sigma^>$ is obtained by the interchange 
$>\longleftrightarrow <$. The resulting kinetic equation reads
\begin{eqnarray}
&&{\partial f_1\over\partial t}+{\partial\varepsilon_1\over\partial k}
{\partial f_1\over\partial r}-{\partial\varepsilon_1\over\partial r}
{\partial f_1\over\partial k}
\nonumber\\
&&=\int {dpdq\over(2\pi)^6}
2\pi\delta\left(\varepsilon_1+\bar\varepsilon_2-\bar\varepsilon_3-
\bar\varepsilon_4-2\Delta_E\right)
\nonumber\\
&&\times
\Biggl(1-{1\over 2}{\partial\Delta_2\over\partial r}-
{\partial\bar\varepsilon_2\over\partial r}
{\partial\Delta_2\over\partial\omega}\Biggr)
\nonumber\\
&&\times
\left|t\Bigl(\!\varepsilon_1\!+\!\bar\varepsilon_2\!-\!\Delta_E,
k\!-\!{1\over 2}\!\Delta_K,\!p-\!{1\over 2}\!\Delta_K,q,r\!-\!\Delta_r,
t\!-\!{1\over 2}\!\Delta_t\!\Bigr)\!\right|^2
\nonumber\\
&&\times
\left((1-f_1)(1-\bar f_2)\bar f_3\bar f_4-
f_1\bar f_2(1-\bar f_3)(1-\bar f_4)\right).
\nonumber\\
\label{vc12}
\end{eqnarray}
The abbreviated notation of $f$'s and $\varepsilon$'s means
\begin{equation}
\begin{array}{rcl}
\varepsilon_1&\equiv&\varepsilon(k,r,t),\\
\bar\varepsilon_2&\equiv&\varepsilon(p,r-\Delta_2,t),\\
\bar\varepsilon_3&\equiv&\varepsilon(k-q-\Delta_K,r-\Delta_3,t-
\Delta_t),\\
\bar\varepsilon_4&\equiv&\varepsilon(p+q-\Delta_K,r-\Delta_4,t-
\Delta_t).
\end{array}
\label{vc9d}
\end{equation}
The bar reminds that all $\Delta$'s appear in arguments with negative 
signs. In the $\Delta$'s and their derivatives the energy $\omega+E=
\varepsilon_1+\bar\varepsilon_2$ is substituted after all derivatives 
are taken.
Note that the pre-factor has changed compared to (\ref{vc7d}). In (\ref{vc7d}),
$\bar\varepsilon_2$ depends on $\Delta_2$ which depends on the energy $E$. 
This $E$-dependence has resulted in a norm of the singularity.

The non--local kinetic equation (\ref{vc12}) represents the main
result of this paper. In the following we will discuss some symmetry
properties of this equation.

\subsection{Markovian approximation}
In the scattering integrals of (\ref{vc9d}) the collision delay $\Delta_t$ 
appears together with displacements $\Delta_{2-4}$. Actual values of these
types of corrections are closely linked. In the original Wigner's approach,
the collision delay was identified from the position of the center of wave
packets in the final state. Apparently, this position can be described 
either as the space displacement or via the delay. A corresponding 
rearrangement of the scattering integral is achieved using the free-space
time evolution of the distribution, e.g.,
\begin{eqnarray}
\bar f_3&=&f(k-q-\Delta_K,r-\Delta_3,t-\Delta_t)
\nonumber\\
&\approx&f(k-q-\Delta_K+F_3\Delta_t,r-\Delta_3+v_3\Delta_t,t),
\label{frD}
\end{eqnarray}
where $v_3=\partial_k\varepsilon_3$ is a velocity and $F_3=-\partial_r
\varepsilon_3$ is a force acting on the particle in the final state.
Substituting the right hand side (and similar for $\bar f_4$ and 
$\bar\varepsilon_{3,4}$) into 
(\ref{vc9d}), one obtains instantaneous scattering integrals. This liberty 
of rearrangement is only approximative, some quantities are sensitive to
such approximations. For instance, the correlated density discussed below
is directly proportional to $\Delta_t$ as it measures an effective number 
of particles in the collision process. On the other hand, the rearrangement
into the instantaneous form makes the kinetic equation Markovian what
significantly simplifies its numerical treatment, in particular if 
negative collision delay takes place.

\subsection{Space-time symmetry of non-local corrections}
The scattering-out integral in (\ref{vc12}) has the same sign of
$\Delta$'s as the scattering-in. Accordingly, due to the non-local and
non-instantaneous corrections, the scattering-out cannot be obtained
from the scattering-in by the space-time symmetry, as it is common in 
the classical approach to the kinetic equation. This problem has a
practical impact. The simulations of
pressure with non-instant corrections has been discussed by
\cite{DP96}. In Monte-Carlo simulations based on semi-classical
trajectories the space-time symmetry is an unavoidable part of the
scattering integral. To understand internal times of collisions and
non-local corrections, it is thus necessary to clarify the space-time
symmetry.

It is possible to show that the scattering-in and -out are connected
by the space-time symmetry, one has to take into account, however, the
non-local and non-instantaneous corrections to the T-matrix itself.
To this end we employ the optical theorem which can be expressed in two
forms,
\begin{equation}
{\rm Im}{\cal T}={\cal T}^R{\cal A}{\cal T}^A=
{\cal T}^A{\cal A}{\cal T}^R.
\label{vc13}
\end{equation}
Here, symbol Im denotes the anti-Hermitian part. The multiplication
includes integrals over two-particle states and times. The function
${\cal A}_{34}=G^>_3G^>_4-G^<_3G^<_4=G^>(3,3')G^>(4,4')-G^<(3,3')
G^<(4,4')$ is the two-particle
spectral function which includes the Pauli blocking of internal states
in the Galitskii-Feynman approximation. Both forms of ${\rm Im}{\cal T}$
can be derived by algebraic manipulations with the ladder equation for
${\cal T}^{R,A}$ and hold out of equilibrium.

Implementation of the optical theorem (\ref{vc13}) requires to rearrange the
scattering integrals. This is most conveniently done on the level of the
starting non-equilibrium Green functions. The integrand in the right hand 
side of (\ref{levin}) is composed of two terms, in a reduced notation 
$I_l=G^>_1\Sigma^<_1-G^<_1\Sigma^>_1$ and $I_r=\Sigma^<_1G^>_1-\Sigma^>_1
G^<_1$. For the Galitskii-Feynman selfenergy, 
\begin{equation}
I_l=G^>_1G^>_2{\cal T}^RG^<_3G^<_4{\cal T}^A-
G^<_1G^<_2{\cal T}^RG^>_3G^>_4{\cal T}^A,
\label{vc15p}
\end{equation}
we have to add and subtract $G^<_1G^<_2{\cal T}^RG^<_3G^<_4{\cal T}^A$ to the second
term of (\ref{vc15p}) to achieve the Galitskii-Feynman two-particle 
spectral function, 
\begin{equation}
I_l={\cal A}_{12}{\cal T}^RG^<_3G^<_4{\cal T}^A-
G^<_1G^<_2{\cal T}^R{\cal A}_{34}{\cal T}^A.
\label{vc15}
\end{equation}
In the last term we can apply the optical theorem (\ref{vc13}) for the 
interchange ${\cal T}^R{\cal A}_{34}{\cal T}^A\longrightarrow{\cal T}^A
{\cal A}_{34}{\cal T}^R$, therefore
\begin{equation}
I_l={\cal A}_{12}{\cal T}^RG^<_3G^<_4{\cal T}^A-
G^<_1G^<_2{\cal T}^A{\cal A}_{34}{\cal T}^R.
\label{vc15a}
\end{equation}
In the same way one can rearrange $I_r$.

Technically, the interchange of the retarded and the advanced T-matrices
is equivalent to the change of sign of the phase shift $\phi$, ${\cal T}^R
\to t^R=|t|{\rm e}^{-i\phi}$ and ${\cal T}^A\to t^A=|t|{\rm e}^{i\phi}$.
The implementation of the optical theorem thus has no effect on the local
parts of the scattering integral which depend exclusively on the amplitude
$|t|$ of the T-matrix. In contrast, all non-local corrections given by
$\Delta$'s (\ref{vc4a}-\ref{vc4g}) are linear functions of the phase shift
$\phi$ and thus reverse their signs. In the kinetic equation (\ref{vc12}),
the combination corresponding to ${\cal A}_{34}$ is achieved by regrouping 
distributions as
\begin{eqnarray}
&&(1-f_1)(1-\bar f_2)\bar f_3\bar f_4-
f_1\bar f_2(1-\bar f_3)(1-\bar f_4)
\nonumber\\
&&=(1-f_1-\bar f_2)\bar f_3\bar f_4-
f_1\bar f_2(1-\bar f_3-\bar f_4).
\label{vc16}
\end{eqnarray}
The intermediate-state Pauli-blocking factor $1-\bar f_3-\bar f_4$ is
the occupation number corresponding to ${\cal A}_{34}$. Inverting signs
of the phase shift, i.e., of $\Delta$'s, related to the scattering
integral with the Pauli blocking $1-\bar f_3-\bar f_4$ one arrives at
the kinetic equation
\begin{eqnarray}
&&{\partial f_1\over\partial t}+{\partial\varepsilon_1\over\partial k}
{\partial f_1\over\partial r}-{\partial\varepsilon_1\over\partial r}
{\partial f_1\over\partial k}
\nonumber\\
&&=\int {dp\over(2\pi)^3}{dq\over(2\pi)^3}
\Biggl(1-{1\over 2}{\partial\Delta_2\over\partial r}-
{\partial\varepsilon_2\over\partial r}
{\partial\Delta_2\over\partial\omega}\Biggr)
\nonumber\\
&&\times \left|t\Bigl(\!\varepsilon_1\!+\!\bar\varepsilon_2\!-\!
\Delta_E,k\!-\!{1\over 2}\!\Delta_K,\!p-\!{1\over 2}\!\Delta_K,q,r\!-\!
\Delta_r,t\!-\!{1\over 2}\!\Delta_t\!\Bigr)\!\right|^2
\nonumber\\
&&\times 2\pi\delta\left(\varepsilon_1+\bar\varepsilon_2-\bar\varepsilon_3-
\bar\varepsilon_4-2\Delta_E\right)(1-f_1-\bar f_2)\bar f_3\bar f_4
\nonumber\\
&&-\int {dp\over(2\pi)^3}{dq\over(2\pi)^3}
\Biggl(1+{1\over 2}{\partial\Delta_2\over\partial r}
+{\partial\varepsilon_2\over\partial r}
{\partial\Delta_2\over\partial\omega}\Biggr)
\nonumber\\
&&\times \left|t\Bigl(\!\varepsilon_1\!+\!\varepsilon_2\!+\!\Delta_E,
k\!+\!{1\over 2}\!\Delta_K,\!p+\!{1\over 2}\!\Delta_K,q,r\!+\!\Delta_r,
t\!+\!{1\over 2}\!\Delta_t\!\Bigr)\!\right|^2
\nonumber\\
&&\times 2\pi\delta\left(\varepsilon_1+\varepsilon_2-\varepsilon_3-
\varepsilon_4+2\Delta_E\right)f_1f_2(1-f_3-f_4).
\nonumber\\
\label{vc17}
\end{eqnarray}
In functions with bars arguments are given by (\ref{vc9d}), in functions
without bars arguments are identical except for the reversed (i.e., 
positive) signs of all $\Delta$'s.

The kinetic equation with the space-time symmetry of scattering integrals
reminds to the Enskog equation for classical hard spheres with respect
to two features. 
First, if one suppresses the collision delay, see (\ref{frD}), the 
scattering-out becomes the space inversion of the scattering-in. Second,
the scattering rate (given by $|t|^2$) is centered between initial and
final states of collisions. 

The kinetic equation (\ref{vc17}) has a difficult interpretation because its 
scattering integrals have positive or negative values following the sign of
the Pauli blocking factors $1-f_1-f_2=(1-f_1)(1-f_2)-f_1f_2$ and $1-f_3-f_4=
(1-f_3)(1-f_4)-f_3f_4$. This strange behavior appears due to the stimulated 
transition processes given by the terms $-f_1f_2$ and $-f_3f_4$ on top of the
standard Pauli blocking, $(1-f_1)(1-f_2)$ and $(1-f_3)(1-f_4)$. While in
the local approximation of the scattering integrals the -in and -out 
stimulated transitions exactly compensate giving no net contribution, the
non-local corrections translate this feature known from the equilibrium
Green functions on the level of kinetic equation. Our difficulties with
the classical interpretation of (\ref{vc17}) thus may follow from our low
understanding of stimulated processes in general. At very low temperatures, 
the stimulated transitions become so strong that the above quasiparticle 
picture breaks down and the system develops super-conductivity having an
additional degree of freedom and a different spectrum of single-particle 
excitations. Although the Galitskii-Feynman approximation is not suited to 
describe the super-conducting state, it reveals a singularity at $T_c$ which 
signals its onset. 

Let us put aside problems with stimulated processes and return to our 
discussion of characteristic times. For the moment we will assume that 
the collision delay $\Delta_t$ is positive. In the scattering-in integral 
which is the first term on the right hand side of (\ref{vc17}), the 
initial condition is at the time $t-\Delta_t$, i.e., before the time $t$ at 
which we balance the changes at the phase-space point $(k,r)$. This has a clear 
interpretation, at $t-\Delta_t$ the particle of momentum $k-q$ enters
into the collision process from which it is released at $t$ into $(k,r)$.
In the scattering-out process, the particle leaves the cell $(k,r)$ at 
$t$ when it enters the collision process. This process will end at 
$t+\Delta_t$ when we have to check the accessible phase space. While the 
scattering-out has a natural interpretation, its description violates
the causality. Indeed, the occupation factors in future time $t+\Delta_t$
are supposed to decide whether the process comes through or whether it is blocked.
We note that the violation of the causality on the microscopic time scale
is common in kinetic equations. Already the Fermi Golden Rule includes
a time integral into future which eliminates off-shell processes. In fact,
it is the Fermi-Golden-Rule-type approximation, in this paper represented
by the pole approximation, which makes negative values of the
collision delay possible. 

For a detailed discussion of the space-time
versus particle-hole symmetry for various forms of the two--particle
spectral function see \cite{SLM98}.

\subsection{Correlated density}
As mentioned in the introduction, in the system described on the level of
the Galitskii-Feynman approximation the density of quasiparticles, $n_f=
\int{dk\over(2\pi)^3}f$, differs from the density of composing particles 
$n=\int{dk\over(2\pi)^3}\rho$ by the amount called the correlated density 
$n_c=n-n_f$. After some algebra, from relation (\ref{an1}) one finds
\begin{eqnarray}
n_c&=&\int{dkdpdq\over(2\pi)^8}|t|^2\Delta_t
\nonumber\\
&&\times\delta(\varepsilon_1+
\varepsilon_2-\varepsilon_3-\varepsilon_4)f_1f_2(1-f_3-f_4).
\label{nc1}
\end{eqnarray}
The functions $|t|$, $\varepsilon$'s and $f$'s are in the local approximation 
having no $\Delta$'s in arguments. This expression is the generalized
Beth-Uhlenbeck formula \cite{BKKS96,SZ79,SR87,BU37,ZS85,SRS90,MR95}. The 
correlated density is proportional to the energy-derivative of the phase 
shift expressed here via the collision delay.

This correlated density is consistent with the equation of continuity 
found from the kinetic equation (\ref{vc17}). For simplicity we will assume 
a homogeneous system for the equation of continuity reading, $\partial_t n
=0$. Taking the momentum integral over (\ref{vc17}) one finds
\begin{eqnarray}
{\partial\over\partial t}&&\int{dk\over(2\pi)^3}f_1=-
{\partial\over\partial t}\int{dkdpdq\over(2\pi)^8}|t|^2\Delta_t
\nonumber\\
&&\times\delta(\varepsilon_1+
\varepsilon_2-\varepsilon_3-\varepsilon_4)f_1f_2(1-f_3-f_4).
\label{nc2}
\end{eqnarray}
Using (\ref{nc1}) in the right hand side, one finds $\partial_t(n_f+n_c)=0$.
It should be noted that the number of quasiparticles is not conserved,
$\partial_t n_f\not=0$, because during the time interval $\Delta_t$ the 
colliding particles are excluded from the single-particle statistics being
in two-particle scattering states. 

One can see that instantaneous approximations of the kinetic equation 
cannot capture the correlated density. In the same time, the correct 
description of the quasiparticle density makes the choice (\ref{vc4d}) of the 
collision delay preferable to other characteristic times one can define in
the spirit of Wigner from the asymptotic behavior of final states of binary 
collisions. In Appendix~\ref{secnc} we comment on some mistakes done in
previous studies of the correlated density.

The complete proof of conservation laws with explicit expressions for
the correlated pressure and energies can be found in \cite{LSM97}.

\section{Summary}
In absence of the time operator, there are many definitions of 
characteristic times one can associate with collisions. Regardless of 
a definition one uses, if the finite duration of collisions is included
in a kinetic equation, this equation is non-Markovian since the initial
and final states of the collision are at distinct times. The
family of non-Markovian kinetic equation also includes retarded 
equations of Levinson type in which the collision is expressed in terms 
of the time integral describing the whole process of the two-particle 
interaction.

For slowly varying systems, we have shown that one half of the Levinson-type 
retardation describes the off-shell motion and corresponding 
renormalizations of single-particle functions. The second half compensates
the decay of propagators during the integration and can be eliminated 
leaving the pole contribution to the scattering integral identical to the
one obtained from the quasiparticle approximation.

Finally, we have discussed the collision delay resulting from the energy
dependence of the scattering phase shift. We have shown that this 
non-Markovian correction is connected with other non-local corrections
to the scattering integral which gives us a freedom to define the collision
duration in different ways. 
The physical quantity sensitive to the actual choice of the collision
delay is the density of quasiparticles, or its complementary quantity,
the density of correlated particles. The method presented in this paper,
and at the same time the choice of the collision delay, reproduces the 
correlated density obtained within the generalized Beth-Uhlenbeck approach.

The non-instant and non-local corrections given by the $\Delta$'s do not
change the structure and the overall interpretation of the scattering
integral but only slightly renormalize its ingredients. The exclusive
dependence of the non-local and non-instant corrections on the
scattering phase shift confirms results from the theory of gases
\cite{TNL89,NTL89,H90,NTL91} obtained by very different technical tools.
The non-instant and non-local scattering integral in the form (\ref{vc17})
parallels the classical Enskog's equation, therefore it can be treated
with numerical tools developed for the theory of classical gases, see
e.g. \cite{AGA95} at least as long as $1-f_1-f_2$ remains positive.

The virial corrections to the balance equations appear from intrinsic gradients instead of
correlation parts in the equation of the reduced density matrix. 
In \cite{LSM97} it is shown that the nonlocal kinetic equation
leads to complete balance equations for the density, energy and stress
tensor
which establish conservation laws including correlated parts. The
nonlocal kinetic equation derived here requires no more numerical
efforts than solving the Boltzmann equation. The numerical solution
and application has been demonstrated in \cite{MLSCN98,MT00,MTP00}.

\acknowledgements
K.M. likes to thank the LPC for a friendly and hospitable
atmosphere and Rainer Klages for discussions and useful hints.

\appendix

\section{T-matrix approximation}\label{at}
In order to describe short-ranged two-particle interactions, it is 
necessary to introduce the standard approximation of the many-body theory, 
the T-matrix approximation \protect\cite{KB62,D84,d90}. In the following 
we give a brief compilation of important formulas.

The sum of ladder diagrams is defined as the causal T-matrix
\begin{equation}
\langle 12|{\cal T}|1'2'\rangle\!=\!{\cal V}(121'2')\!+\!{\cal V}(1233')
G(3{\bar 1})G(3'{\bar 2})\langle {\bar 1}{\bar 2}|{\cal T}|1'2'\rangle .
\label{Tmatrix}
\end{equation}
Arguments not present on the left hand side are integrated over. For 
simplicity, we do not assume the sum over spin and isospin and the 
anti-symmetrization for identical particles. Since the potential is local 
in time, we can simplify the T-matrix as $\langle 12|{\cal T}|1'2'\rangle = 
\langle x_1x_2t_1|{\cal T}|x_1'x_2't'_1\rangle \delta(t_1-t_2)\delta
(t_1'-t_2')$, and ladder summation (\protect\ref{Tmatrix}) reads
\begin{eqnarray}
&&\langle x_1x_2t|{\cal T}|x_1'x_2't'\rangle 
={\cal V}(x_1x_2x_1'x_2')\delta(t-t')
\nonumber \\
&&+{\cal V}(x_1x_2x_3x_3')\langle x_3x_3't|{\cal G}|{\bar x_1}{\bar
x_2}{\bar t}\rangle 
\langle {\bar x_1}{\bar x_2}{\bar t}|{\cal T}|x_1'x_2't'\rangle ,
\label{ladders}
\end{eqnarray}
where we have introduced the two-particle Green function,
\begin{equation}
\langle x_1x_2t|{\cal G}|{\bar x_1}{\bar x_2}{\bar t}\rangle =
G(x_1t{\bar x_1}{\bar t})G(x_2t{\bar x_2}{\bar t}).
\label{abbreviation}
\end{equation}
The causal selfenergy then reads
\begin{equation}
\Sigma(11')=-i\int dx_2dx_2'G(x_2't_1'x_2t_1^+)
\langle x_1x_2t_1|{\cal T}|x_1'x_2't_1'\rangle .
\label{tmatrix}
\end{equation}

Using the Langreth-Wilkins rules \cite{LW72} (which are equivalent to 
the algebra on the Keldysh contour) we obtain the real-time Green's 
functions. From (\ref{tmatrix}) follows
\begin{eqnarray}
\Sigma^{\gtrless}&=&{\cal T}^{\gtrless}G^{\lessgtr} 
\nonumber\\
\Sigma^{R/A}&=&-i\Theta(t_1-t_2)[{\cal T}^>G^<+{\cal T}^<G^>]
\nonumber\\
&=&{\cal T}^{R}G^<-{\cal T}^<G^{A}.
\label{selfrel}
\end{eqnarray}
We have abbreviated the notation, all integrations and variables are the 
same as in (\protect\ref{tmatrix}). From (\ref{ladders}) follows
\protect\cite{bm90,D84,KSB85}
\begin{eqnarray}
{\cal T}^{\gtrless}&=&{\cal T}^R{\cal G}^{\gtrless} {\cal T}^A 
\label{bt31}\\
{\cal T}^{R/A}&=&{\cal V}+{\cal V}{\cal G}^{R/A}{\cal T}^{R/A}.
\label{optical}
\end{eqnarray}
Explicitly (\ref{bt31}) reads
\begin{eqnarray}\label{opticaltheorem}
&&<x_1x_2t|{\cal T}^{\gtrless}_{ab}|x_1'x_2't'> 
\nonumber\\
&& =  \int d{\bar x_1}d{\bar x_1'}d{\bar
x_2}d{\bar x_2'}
d{\bar t}d{\bar t'} <{x_1}{x_2}t|{\cal T}^R_{ab}|{\bar x_1}{\bar x_2}{\bar t}>
\nonumber\\
&&\times < { \bar x_1}{\bar x_2}{\bar t}|{\cal G}^{\gtrless}_{ab}|{\bar x_1'}{\bar
x_2'}{\bar t'}>
<{\bar x_1'}{\bar x_2'}{\bar t'}|{\cal T}^A_{ab}|x_1'x_2't'>.
\end{eqnarray}

In the scattering integral we use the following representation of the 
two-particle T-matrix,
\begin{eqnarray}
&&\langle x_1x_2\bar t|{\cal T}|x_1'x_2't'\rangle 
=\int\frac{dkdpdqd\Omega}{(2\pi)^{10}} 
{\cal T}\left(\Omega,k,p,q,r,t\right)
\nonumber\\
&&\times\exp\left[i\left (kx_1+px_2-(k-q)x_1'-(p+q)x_2'-\Omega(\bar
    t-t')\right )\right],
\nonumber\\
\label{fourier}
\end{eqnarray}
where $r={1\over4}(x_1+x_2+x_1'+x_2')$ and $t={1\over 2}(\bar t+t')$ are
center-of-the-mass coordinate and time. 

\subsection{Quasiparticle approximation}
We introduce the notation for the known quasiparticle approximation 
without gradient corrections. In the quasiparticle approximation the 
free two-particle propagator ${\cal G}$ (denoted in the mixed Wigner
representation as $\zeta$) takes the form
\begin{equation}
\zeta^R(\Omega, p,p',r,t)={1-f(p,r,t)-f(p',r,t)\over\Omega-
\varepsilon(p,r,t)-\varepsilon(p',r,t)+i\eta}.
\label{two}
\end{equation}
The selfenergy then reads
\begin{equation}
\sigma^>(\omega,k,r,t)=\int{dp\over(2\pi)^3}\,
t^>(\omega+\varepsilon(p,r,t),k,p,0)f(p,r,t),
\label{sig}
\end{equation}
and analogously $\sigma^<$. In the rest of this chapter we suppress 
variables $r,t$. From (\ref{optical}) one finds
\begin{eqnarray}
t^{<}(\Omega,k,p,0)&=&\int{dq\over(2\pi)^3}
|t^R|^2(\varepsilon_{k-q}+\varepsilon_{p+q},k,p,q)
\nonumber\\
&\times&  f_{k-q}f_{p+q} 2 \pi \delta(\omega-\varepsilon_{p+q}-
\varepsilon_{k-q}),
\label{opti}
\end{eqnarray}
and ${t}^{>}$ is given by interchange $f\leftrightarrow 1-f$.

For the determination of the quasiparticle energy one needs the
retarded selfenergy 
\begin{eqnarray}
\sigma^R(\omega,k)&=&-\!\int \!\!\!{dp\over(2\pi)^3}{d\omega'\over 2\pi}
{t^>(\omega'+\varepsilon_p,k,p,0)f_p\over\omega'-\omega -i\eta}
\nonumber\\ 
&&-\!\int \!\!\!{dp\over(2\pi)^3}{d\omega'\over 2\pi}
{t^<(\omega'+\varepsilon_p,k,p,0)(1-f_p)\over\omega'-\omega-i\eta}
\nonumber\\ 
&=&\int{dp\over(2\pi)^3}\,t^R(\omega+\varepsilon_p,k,p,0)f_p 
\nonumber\\
&-&\!\int\!\!{dpdq\over(2\pi)^6}{|t^R(\varepsilon_{k-q}+\varepsilon_{p+q},
k,p,q)|^2f_{k-q}f_{p+q}\over\varepsilon_{k-q}+\varepsilon_{p+q}-
\varepsilon_p-\omega+i\eta}.
\nonumber\\
\label{selfe}
\end{eqnarray}
In the real part of the selfenergy the real part of the T-matrix appears,
\begin{eqnarray}
{\rm Re} t^R(\Omega, k,p,0)&=&\int{dq\over(2\pi)^3}
{1-f_{k-q}-f_{p+q}\over\Omega-\epsilon_{k-q}-\epsilon_{p+q}}
\nonumber\\
&\times&|t^R|^2(\epsilon_{k-q}+\epsilon_{p+q},k,p,q).
\eeq
This form follows from the Kramers-Kronig transformation of $t^>-t^<$. 

The ladder equation without gradients reads [notation is specified
by (\ref{notb2})]
\beq
&&\langle p_1|t^R(q')|p_2\rangle =\langle p_1|{\cal V}(q')|p_2\rangle +
\!\int \!\!\!{dp'dp''\over(2\pi)^6}
\nonumber\\
&&\times
\langle p_1|{\cal V}(q')|p'\rangle \langle p'|\zeta^R(q')|p''\rangle 
\langle p''|t^R_{sc}(q')|p_2\rangle .
\label{bsd}
\eeq
In the coordinates, $k={q'\over 2}+p_1,p={q'\over 2}-p_1,q=p_1-p_2$, 
corresponding to (\ref{fourier}) this equation takes the form
\beq
&&t^R(\Omega,k,p,q)={\cal V}(k,p,q)\nonumber\\
&&+\int{dp'dp''\over(2\pi)^6}
{\cal V}(k,p,p')\zeta^R(\Omega,k-p',p+p',p'-p'')
\nonumber\\
&&\times t^R(\Omega,k-p'',p+p'',q+p'').
\nonumber\\&&
\eeq
Since $\zeta^{\gtrless}(k,p,q)=(2\pi)^3\delta(q)g^{\gtrless}(k)
g^{\gtrless}(p)$, in the quasiparticle approximation one obtains
\beq\label{bs}
&&t^R(\Omega,k,p,q)={\cal V}(k,p,q)+\int{dq'\over(2\pi)^3}{\cal V}(k,p,q')
\nonumber\\
&\times&{1-f_{k-q'}-f_{p+q'}\over\Omega-\varepsilon_{p+q'}-
\varepsilon_{k-q'}+i\eta}t^R(\Omega,k-q',p+q',q+q').
\nonumber\\
\eeq

\section{Gradient expansion}
\label{expansion}
Let us consider a product of two-particle functions
\begin{equation}
{\cal C}(1234)=\int d3'd4'{\cal A}(123'4'){\cal B}(3'4'34).
\label{int}
\end{equation}
We transform this product into mixed representation (\ref{fourier})
keeping gradients till the linear order. 

Variables in (\ref{fourier}) follow variables of the scattering integral. 
For the scattering-in, the momenta correspond to the process $k-q,p+q\to 
k,p$. These variables, however, do not form a convenient algebra for the 
gradient expansion. To this end we use the coordinates 
\be
\alpha&=&1-2
\nonumber\\
\beta&=&3-4
\nonumber\\
\tau&=&{1\over 2}(1+2-3-4)
\nonumber\\
x&=&{1\over 4}(1+2+3+4)\equiv(r,t).
\label{notb2}
\ee 
Since the relative coordinates, $\alpha$ and $\beta$, obey the standard matrix 
algebra while only the coordinate $\tau$ and $x$ undergo the gradient 
expansion, we write these arguments in form $\langle\alpha|C(\tau,x)|\beta
\rangle$. In this notation the product (\ref{int}) reads
\beq
\langle\alpha|C(\tau,x)|\beta\rangle
&=&\int d\gamma d\bar\tau\langle\alpha|A\left(\tau-\bar\tau,x+{1\over 2}
\bar\tau\right)|\gamma\rangle 
\nonumber\\
&\times&\langle\gamma|B\left(\bar\tau,x+{1\over 2}(\tau-\bar\tau)\right)
|\beta\rangle .
\eeq
Assuming a slow variation of the center-of-mass coordinate $x$ we can
write this product keeping gradients in $x$ up to the linear order. The 
relative coordinates can be suppressed being independent of this expansion.
Via the Fourier transformation of coordinate $\tau$,
\begin{equation}
C(\tau,x)=\int{d\kappa\over(2\pi)^4}c(\kappa,x) 
{\rm e}^{i\sum_{n=1}^3\kappa_n\tau_n-i\kappa_4\tau_4},
\label{ftO}
\end{equation}
we express the operator ${\cal C}$ as a function of the sum momentum,
$(\kappa_1,\kappa_2,\kappa_3)=k+p$, and energy $\kappa_4=\Omega$ of both 
particles. One can see that the product has the same form as products of 
single-particle functions,
\begin{equation}
c=ab-{i\over 2}[a,b]=ab\left(1-{i\over 2}[\ln a,\ln b]\right),
\label{gradexp}
\end{equation}
where the (generalized) Poisson bracket applies to the sum coordinate 
and time. The logarithmic form of (\ref{gradexp}) is merely a convenient 
notation because $a$ and $b$ are matrices in $\alpha$ and $\beta$. The
logarithm is taken of each matrix element of $a$ and $b$ and summed
together with the product $ab$.

In parallel with the Dyson equation, there are no linear gradients in the
ladder equation (\ref{optical}). Like the retarded Green
function, the T-matrix is given by the matrix inversion, ${\cal T}_R^{-1}=
{\cal V}^{-1}-{\cal G}^R$. Since the T-matrix is symmetric with respect to
matrix arguments, $\langle s|{\cal T}|s'\rangle=\langle s'|{\cal T}| 
s\rangle$ ($s$ and $s'$ are momenta associated with $\alpha$ and $\beta$, 
respectively), the matrix inversion does not bring any gradients.

The selfenergy, $\Sigma^<(13)={\cal T}^<(1234)G^>(42)$, includes a number
of gradient contributions due to internal gradients in ${\cal T}^<$ and a
simple non-local correction due the convolution with $G^>$. 

\subsection{Two-particle matrix products} 
For the calculation of the selfenergy in (\ref{vc2}) we need $t^<$ for 
the zero transferred momentum, $q=0$, and its infinitesimal vicinity. 
Making the gradient expansion of (\ref{bt31}) in terms of the Poisson 
brackets one finds
\begin{eqnarray}
t^<&=&t^R\tilde g^<t^A\Biggl(1-{i\over 2}\Bigl(
[\ln t^R,\ln\tilde g^<]-[\ln t^A ,\ln\tilde g^<]
\nonumber\\
&&\ \ \ \ \ \ \ \ \ \ \ 
+[\ln t^R,\ln t^A]\Bigr)\Biggr)
\nonumber\\
&=&|t|^2\tilde g^<\Bigl(1-[\phi,\ln\tilde g^<]-
[\phi,\ln|t|]\Bigr).
\label{tg}
\end{eqnarray}
In the second line we have decomposed the T-matrices into amplitudes and
phase shifts, $t^{R,A}=|t|{\rm e}^{\mp i\phi}$. Note that all gradient 
corrections depend exclusively on the derivatives of phase shift $\phi$. 
Using definitions (\ref{vc4a}-\ref{vc4g}) and the linear approximation,
$a(x)(1-\Delta\partial_x a)=a(x-\Delta)$, expression (\ref{tg}) can be 
given a form
\begin{eqnarray}\label{result1}
&&t^<\left(\Omega,k,p,0,r-{1\over 2}\Delta_2,t\right)
=\int {dq\over(2\pi)^3}{dQ\over(2\pi)^3}
\nonumber\\
&&\times t^R\left(\Omega-\Delta_\omega,k-{\Delta_k \over 2},
p-{\Delta_k\over 2},q,r-\Delta_r,t-{\Delta_t\over 2}\right)
\nonumber\\
&&\times\zeta^<\!\left(\!\Omega\!-\!2\Delta_\omega,k\!-\!q\!-\!\Delta_k,
p\!+\!q\!-\!\Delta_k,Q,r\!-\!{\Delta_3\over 2}\!-\!{\Delta_4\over 2},t\!-\!
\Delta_t\!\right)
\nonumber\\
&&\times t^A\left(\Omega-\Delta_\omega,k-{\Delta_k \over 2},p-
{\Delta_k\over 2},q+{Q},r-\Delta_r,t-{\Delta_t\over 2}\right).
\nonumber\\
\label{tls}
\end{eqnarray}
Factors of half in the non-local corrections to the amplitude $|t^{R,A}|$
result from the fact that this amplitude enters the collision integral
in the square.

In the absence of gradients, the two-particle function $\zeta^<$ is
singular in the `transferred' momentum $Q$ giving the only contribution
for $Q=0$. When the gradient corrections are already in the explicit 
form, one can employ this symmetry as it has been used above. Indeed, 
the complex conjugacy of the advanced and retarded T-matrices, $t^A=\bar 
t^R$, requires equal arguments of both functions. Now we show that the
infinitesimal vicinity of $Q=0$ brings additional gradient contributions.

\subsection{Convolution of initial states}
The two-particle correlation function $\zeta^<$ representing initial
states of the collision in (\ref{tls}) is not a suitable input for the
kinetic equation. We have to express $\zeta^<$ in terms of the 
convolution of two single-particle functions. This convolution brings
gradient corrections by which effective positions of particles entering
the collision process become distinct.

>From the inverse transformation to (\ref{fourier}) we obtain
\begin{eqnarray}
&&\zeta^<(\Omega,k-p,p+q,Q,r,t)
\nonumber\\
&&=2^3\int dx_1dx_2dx_1'dx_2'd\bar tdt'\exp[iQ(x_1'-x_2')+i\Omega(\bar t-t')]
\nonumber\\
&&\times\exp\left[-i(k-q)(x_1-x_1')-i(p+q)(x_2-x_2')\right]
\nonumber\\
&&\times
\delta(x_1+x_2+x_1'+x_2'-4r)\delta(\bar t+t'-2t)
\nonumber\\
&&\times
G^<(x_1\bar t,x_1't')G^<(x_2\bar t,x_2't').
\label{invft}
\end{eqnarray}
The substitution, $x_1=r+\alpha+\beta_1/2$, $x_1'=r+\alpha-\beta_1/2$,
$x_2=r-\tilde \alpha+\beta_2/2$, $x_2'=r-\tilde \alpha-\beta_2/2$, $\bar t=t+\tau/2$,
$t'=t-\tilde \tau/2$, shows that the $\delta$-functions yield $\tilde
\tau=\tau$ and $\tilde \alpha=\alpha$ and we obtain 
\begin{eqnarray}
&&\zeta^<(\Omega,k-p,p+q,Q,r,t)
=2^3\!\!\!\int d\beta_1d\beta_2d\alpha d\tau\exp[2iQ\alpha]
\nonumber\\
&&\times\exp\left[-i(k-q+Q/2)\beta_1-i(p+q-Q/2)\beta_2+i\Omega\tau\right]
\nonumber\\
&&\times\hat g^<(\beta_1,r+\alpha,t,\tau)\hat g^<(\beta_2,r-\alpha,t,\tau),
\label{sub1}
\end{eqnarray}
where the representation (\ref{sW}) has been used.

Now we linearize the center-of-mass dependence of $g^<$'s in
$\alpha$. The factor $\alpha$ can be represented by a derivative in front
of the integral, $\alpha={i\over 2}\partial_Q-{i\over 4}\partial_q$.
The remaining integration over $\alpha$ results in $\pi^3\delta(Q)$. When 
substituted into (\ref{tls}), the differential operator in front of the 
integral is treated with the help of the integration by parts giving 
\be
\alpha\to-{1\over 2}(\Delta_3-\Delta_4)
\ee 
and $Q$ is integrated out. In other words the $\alpha$ in the
arguments of $\hat g$ in (\ref{sub1})
can be replaced by nonlocal shifts valid up to linear orders in
gradients. This leads with (\ref{tls}) to
\begin{eqnarray}
&&t^<\left(\Omega,k,p,0,r-{1\over 2}\Delta_2,t\right)
=\int {dq\over(2\pi)^3}
\nonumber\\
&&\times\left|t^R\left(\Omega-\Delta_\omega,k-{\Delta_k \over 2},
p-{\Delta_k\over 2},q,r-\Delta_r,t-{\Delta_t\over 2}\right)\right|^2
\nonumber\\
&&\times
g^<(k-q,r-\Delta_3,t-\Delta_t,\Omega-2 \Delta_\omega) 
\nonumber\\
&&\times
g^<(p+q,r-\Delta_4,t-\Delta_t,\Omega-2 \Delta_\omega) 
.
\nonumber\\
\label{tls1}
\end{eqnarray}

\subsection{Convolution of T-matrix and hole Green function}
Next we evaluate the convolution of the ${\cal T}$-matrix with the
hole Green function which is required for the selfenergy. Writing the selfenergy in the above 
representation, [$\tau=1-3$ and $x={1\over 2}(1+3)$]
\beq
\Sigma^<(\tau,x)&=&\int d\alpha d\beta 
G^>\left(-\tau-\beta+\alpha,x-{1\over 2}(\alpha+\beta)\right)
\nonumber\\
&\times&\langle\alpha|{\cal T}^<\left(\tau-{1\over 2}
(\alpha-\beta),x-{1\over 4}(\alpha+\beta)\right)|\beta\rangle ,
\nonumber\\
\label{selal}
\eeq
we see that the convolution couples matrix arguments $\alpha$ and $\beta$
with the center-of-mass variables. By substitution, $\lambda={1\over 2}
(\alpha+\beta)$ and $\mu=\alpha-\beta$, and expansion in gradients we obtain
\beq
\Sigma^<(\tau,x)&&=\int d\mu\ G^>\left(\mu-\tau,x\right)
\nonumber\\
&&\times\int d\lambda\left\langle\lambda+{\mu\over 2}\left|{\cal T}^<
\left(\tau-{\mu\over 2},x\right)\right|\lambda-{\mu\over 2}\right\rangle 
\nonumber\\
&&-{1\over 2}\int d\mu\ G^>\left(\mu-\tau,x\right)
\nonumber\\
&&\times{\partial\over\partial x}\int d\lambda\ \lambda
\left\langle\lambda+{\mu\over 2}\left|{\cal T}^<\left(\tau-{\mu\over 2},
x\right)\right|\lambda-{\mu\over 2}\right\rangle 
\nonumber\\
&&-\int d\mu\left({\partial\over\partial x}G^>\left(\mu-\tau,x\right)\right)
\nonumber\\
&&\times\int d\lambda\ \lambda\left\langle\lambda+{\mu\over 2}\left|
{\cal T}^<\left(\tau-{\mu\over 2},x\right)\right|\lambda-{\mu\over 2}
\right\rangle .
\nonumber\\
\label{selalp}
\eeq
The second and third terms are gradient corrections due to the convolution.

Now we can transform the selfenergy (\ref{selalp}) into the mixed representation. 
The momentum representation of the matrix algebra is introduced via unity 
operators, e.g., $1=\int{ds\over(2\pi)^3}|s\rangle\langle s|$, with
$\langle\lambda-\mu/2|s\rangle={\rm e}^{is(\lambda-\mu/2)}$ and
$\langle s'|\lambda+\mu/2\rangle={\rm e}^{-is'(\lambda+\mu/2)}$. For the
non-gradient part the integration over $\lambda$ results in the known
fact that only the diagonal element, $s=s'$, contributes. For the gradient
contribution, $\lambda\to -(i/2)(\partial_s-\partial_{s'})$, and the
diagonal element is taken after the derivatives are performed. Since
we have considered the   
gradient corrections to the ${\cal T}$-matrix already in the last
chapter and since the shifts are additive, it is sufficient here to use the 
zero-order approximation 
of $t^<$,
\begin{equation}
\langle s|t^<|s'\rangle=\int{d\bar sd\tilde s\over(2\pi)^2}
\langle s|t^R|\bar s\rangle\langle\bar s|\zeta^<|\tilde s\rangle
\langle\tilde s|t^A|s'\rangle,
\label{slpp}
\end{equation}
where all functions have the center-of-mass argument $(\kappa,x)$. From
$t^{R,A}=|t|{\rm e}^{\mp i\phi}$ and condition $s=s'$ we find that
$\lambda\to -\partial_s\phi$. By the substitution into variables of
the kinetic equation, $(q,0)=s-s'$, $(k,\omega)=\kappa/2+s$, $(p,0)=
\kappa/2-s$ and $(r,t)=x$, one confirms that $-\partial_s\phi={1\over 2}
\Delta_2$, see (\ref{vc4a}), and the selfenergy reads
\beq
\sigma^<_\omega(k,r,t)&=&\int{dpd\Omega\over(2\pi)^4}
g^>_{\Omega-\omega}\left(p,r-\Delta_2,t\right)\left(1-{1\over 2}
{\partial\Delta_2\over\partial r}\right)
\nonumber\\
&\times& t^<\left(\Omega,k,p,0,r-{1\over 2}\Delta_2,t\right) .
\label{vc2}
\eeq
In this expression, $t^<$ abbreviates the right hand side of (\ref{slpp}),
because the non-local correction $\Delta_2$ is defined only with respect 
to the integral over internal states of the collision. The norm term,
$1-{1\over 2}\partial_r\Delta_2$ results from the interchange, $\partial_r
(\Delta_2 A)=(\partial_r\Delta_2) A+\Delta_2\partial_r A$, one has to make 
before
the derivative is expressed in terms of the displacement, e.g., $g^<(r)-
\Delta_2\partial_r g^<(r)=g^<(r-\Delta_2)$.

Together with (\ref{tls1}) we obtain the result (\ref{vc7d}).

\section{Common met pitfall with the correlated density}\label{secnc}
We note that one of us made a mistake deriving the correlated density 
directly from the retardation in the Levinson-type equation \cite{MR95}. 
The expression found in \cite{MR95} is identical to the correlated density
except for two points, there is a wrong sign and the amplitude is as twice
as large. Since this is a pitfall often met in literature we want to give 
some details here that it can be avoided in future. 

The simple expansion of the Levinson equation (\ref{levinson}) up to the 
first order in memory \cite{MR95} yields 
\begin{equation}
{\partial\over\partial t}\rho={\cal I}+{\partial\over\partial t}{\cal R}.
\label{redc}
\end{equation}
To obtain balance equations for the density including virial corrections, 
one has to integrate (\ref{redc}) over momentum $k$, see \cite{MR95,Kl82}. 
Following the arguments developed for the Boltzmann-type equations, in 
\cite{MR95}, the wrong assumption was used that the correlated part of 
the collision integral, $\tilde n_c=\int dk R$, combines with the left 
hand side of (\ref{redc}), $\tilde n_f=\int dk\rho$, to establish the 
density conservation, ${\partial\over\partial t}(\tilde n_f+\tilde n_c)=0$.
The correlated density derived in this way reminds the result known from
equilibrium \cite{SZ79,SR87,BU37,ZS85,SRS90}. Note that in this picture,
the Wigner distribution in left hand side of the Levinson equation is 
treated as the quasiparticle distribution what is a mistake done usually in
density operator studies \cite{Kl82}. The wrong sign follows from this last 
misinterpretation.  By definition the momentum integral over the Wigner 
distribution yields the full density, $\int dk\rho=n$. The conservation law 
thus tells that either $\int dk{R}=0$ or the Levinson equation does not 
conserve the number of particles. 

As we have shown in this paper the Levinson equation contains additional 
gradient terms which exactly compensate the explicit time gradients (used 
to derive $\tilde n_f$). The above double count follows from the double 
count of the off-shell contributions discussed in this paper.

Finally, the $\tau_{R,A}$ retardation was neglected. Since the
correlated density (\ref{nc1}) found from the off-shell contribution to
the Wigner distribution is consistent with the conservation of the number
of particles found from the $\tau_{R,A}$ retardation, one can see that
these two contributions cancel and the collision integral of the Levinson
equation conserves the number of particles, as it should.

%\bibliography{kmsr,kmsr1,kmsr2,kmsr3,kmsr4,kmsr5,kmsr6,kmsr7,delay2,delay3,spin}
%\bibliographystyle{prsty}

\end{document}